\documentclass[prd,onecolumn,notitlepage,eqsecnum,nofootinbib,floatfix]{revtex4-1}
\usepackage{amssymb}
\usepackage{amsmath}
\begin{document}
\allowdisplaybreaks
\title{Gravitational self-force in nonvacuum spacetimes}   
\author{Peter Zimmerman and Eric Poisson} 
\affiliation{Department of Physics, University of Guelph, Guelph, Ontario, N1G 2W1, Canada} 
\date{June 19, 2014} 
\begin{abstract}
The gravitational self-force has thus far been formulated in
background spacetimes for which the metric is a solution to the
Einstein field equations in vacuum. While this formulation is
sufficient to describe the motion of a small object around a black
hole, other applications require a more general formulation that
allows for a nonvacuum background spacetime. We provide a foundation
for such extensions, and carry out a concrete formulation of the
gravitational self-force in two specific cases. In the first we
consider a particle of mass $m$ and scalar charge $q$ moving in a
background spacetime that contains a background scalar field. In the
second we consider a particle of mass $m$ and electric charge $e$
moving in an electrovac spacetime. The self-force incorporates all
couplings between the gravitational perturbations and those of the
scalar or electromagnetic fields. It is expressed as a sum of local
terms involving tensors defined in the background spacetime and
evaluated at the current position of the particle, as well as tail
integrals that depend on the past history of the particle. Because
such an expression is rarely a useful starting point for an explicit
evaluation of the self-force, we also provide covariant expressions
for the singular potentials, expressed as local expansions near the
world line; these can be involved in the construction of effective
extended sources for the regular potentials, or in the computation of
regularization parameters when the self-force is computed as a sum
over spherical-harmonic modes. 
\end{abstract} 
\maketitle

\section{Introduction and overview} 
\label{sec:intro} 

The prospect of measuring low-frequency gravitational waves generated
by a solar-mass compact object spiraling toward a supermassive black
hole has motivated a large effort to describe the motion of such a body
beyond the test-mass approximation. In this treatment
\cite{mino-etal:97, quinn-wald:97}, the body's gravitational influence
is taken into account, and the motion is no longer geodesic in the
background spacetime of the large black hole. The motion is instead
accelerated, the perturbation created by the small body giving rise to
a gravitational self-force. It is this self-force that is responsible
for the body's inspiraling motion, implied by the loss of orbital
energy and angular momentum to gravitational waves. To date the
gravitational self-force was formulated rigorously
\cite{gralla-wald:08, pound:10a, pound:10b}, it was computed and
implemented in orbital evolutions around nonrotating black holes
\cite{warburton-etal:12}, it was implicated in an improved calculation
of the innermost circular orbit of a Kerr black hole
\cite{isoyama-etal:14}, and it was extended to second order in
perturbation theory \cite{detweiler:12, gralla:12, pound:12a,
  pound:12b, pound:14}. The consequences of the gravitational
self-force have been compared to the predictions of high-order
post-Newtonian theory \cite{shah-friedman-whiting:14} and numerical
relativity \cite{letiec-etal:13}. Other achievements of the
gravitational self-force program are reviewed in
Refs.~\cite{barack:09, poisson-pound-vega:11}.  

The gravitational self-force has thus far been formulated for bodies
moving in vacuum spacetimes, and it is indeed an important restriction
of the formulation that the background metric be everywhere a vacuum
solution to the Einstein field equations. For the applications
reviewed in the preceding paragraph, the background spacetime is 
produced by an isolated black hole, and the vacuum formulation of the
gravitational self-force is perfectly adequate. Other applications,
however, may require an extension to nonvacuum spacetimes. For
example, one might wish to compute the self-force acting on a
satellite of a material body, and this would require a formulation
that allows for the presence of matter somewhere in the spacetime.   

Another application that requires such an extension is the elucidation  
of the role played by the self-force in scenarios that aim to produce
a counter-example to cosmic censorship by overcharging a near-extremal   
Reissner-Nordstr\"om black hole. Back in 1999, Hubeny \cite{hubeny:99} 
showed that a charged black hole near the extremal state can absorb a
particle of such charge, mass, and energy that the final configuration
cannot be a black hole, because its charge-to-mass ratio exceeds the
extremal bound. A variation on this theme was explored by Jacobson and
Sotiriou \cite{jacobson-sotiriou:09}, who revealed that a
near-extremal Kerr black hole can absorb a particle and be driven
toward a final state with too much angular momentum to be a black
hole. These works treated the particle as a test particle in the
black-hole spacetime, and it was soon realized that self-force and
radiative effects can play an important role in these overcharging and
overspinning scenarios. In fact, Hubeny incorporated approximate
self-force effects in her original analysis, and Barausse, Cardoso, and
Khanna \cite{barausse-cardoso-khanna:10} took into account the
gravitational radiation emitted by the particle on its way to overspin
a Kerr black hole. 

These partial attempts to incorporate (conservative and radiative)
self-force effects could not rule out all overcharging and
overspinning scenarios. Another partial attempt
\cite{zimmerman-etal:13}, based on a calculation of the
electromagnetic self-force acting on a charged particle falling toward
a Reissner-Nordstr\"om black hole, was more successful: a thorough
sampling of the parameter space failed to produce a single instance of 
an overcharged final state. This work bolstered the case that the
self-force acts as a cosmic censor in these scenarios, but the
analysis was still an incomplete one: the calculation included the
electromagnetic self-force but neglected the gravitational self-force,
and it did not account for the coupling between electromagnetic and
gravitational perturbations in a background Reissner-Nordstr\"om
spacetime. A complete account of self-force effects in overcharging
scenarios will have to overcome these limitations.    

Our purpose in this paper is to provide a foundation for performing
self-force calculations in nonvacuum background spacetimes. (Other
foundational elements have been provided by Gralla \cite{gralla:10,
  gralla:13}.) We have in mind the situation described previously: a
particle of mass $m$ and electric charge $e$ moves in an electrovac
spacetime with metric $g_{\alpha\beta}$ and electromagnetic field
$F_{\alpha\beta}$, solutions to the Einstein-Maxwell equations. The
particle produces a gravitational perturbation $h_{\alpha\beta}$, an
electromagnetic perturbation $f_{\alpha\beta}$, and we wish to
calculate the complete self-force acting on the particle. This
includes terms originating in $h_{\alpha\beta}$ and scaling as $m^2$,
terms originating in $f_{\alpha\beta}$ and scaling as $e^2$, but there
are also terms originating in the coupling between gravitational and
electromagnetic perturbations and scaling as $em$. 

It is this coupling that gives rise to the most challenging aspects of
the formulation. The physical origin of the coupling is easy to
identify. The gravitational perturbation produced by the particle
gives rise to a shift in the background electromagnetic field, and
the electromagnetic perturbation gives rise to a shift in the field's 
energy-momentum tensor, which in turn produces a shift in the
background metric. To first order in perturbation theory, 
$h_{\alpha\beta}$ is the sum of the gravitational perturbation
generated by the particle and the electromagnetic shift in the
background metric, and $f_{\alpha\beta}$ is the sum of the
electromagnetic perturbation sourced by the particle and the
gravitational shift in the background electromagnetic field.  

The coupling is manifested in the field equations satisfied by the
perturbations. To display this it is convenient to introduce a
``trace-reversed'' gravitational perturbation $\gamma_{\alpha\beta} :=
h_{\alpha\beta} - \frac{1}{2} g_{\alpha\beta} g^{\mu\nu} h_{\mu\nu}$,
a vector potential $b_\alpha$ such that $f_{\alpha\beta} =
\nabla_\alpha b_\beta - \nabla_\beta b_\alpha$, and to collect the
perturbations into a ``meta-potential'' $\psi^A :=
\{\gamma^{\alpha\beta}, b^\alpha\}$. Adopting the Lorenz gauge
conditions $\nabla_\beta \gamma^{\alpha\beta} = 0$ and $\nabla_\alpha
b^\alpha = 0$, we find that the perturbation equations take the form
of the coupled wave equations 
\begin{equation} 
\Box \psi^A + M^A_{\ B\mu} \nabla^\mu \psi^B 
+ N^A_{\ B} \psi^B = -4\pi \mu^A, 
\label{eq1} 
\end{equation} 
where $\Box := g^{\alpha\beta} \nabla_\alpha \nabla_\beta$ is the
covariant wave operator in the background spacetime, $M^A_{\ B\mu}$
and $N^A_{\ B}$ are tensors in the background spacetime, and $\mu^A$
collects the source terms --- the particle's energy-momentum tensor
and current density. Summation over a repeated meta-index $B$ is
understood, and the coupling between gravitational and electromagnetic
perturbations is revealed by the fact that $M^A_{\ B\mu}$ and
$N^A_{\ B}$ possess off-diagonal components --- terms that involve
either a pair of tensorial indices for $A$ and a single tensorial index for
$B$, or a single index for $A$ and a pair of indices for $B$.   

While the coupling introduces a level of complexity not experienced 
before in self-force calculations, the perturbation equations are
of the same mathematical type --- hyperbolic equations --- as those
encountered in previous formulations. As such they can be handled in
exactly the same way, and the starting point is to integrate
Eq.~(\ref{eq1}) with the help of a retarded Green's function 
$G^A_{\ B'}(x,x')$ that satisfies a special case of Eq.~(\ref{eq1})
with a Dirac distribution on the right-hand side. The solution is  
\begin{equation} 
\psi^A(x) = \int G^A_{\ B'}(x,x') \mu^{B'}(x')\, dV', 
\label{eq2} 
\end{equation} 
where $dV'$ is the invariant volume element in the background
spacetime. The Green's function possesses both diagonal and
off-diagonal components, and the right-hand side of Eq.~(\ref{eq2}) 
includes a contribution from the particle's energy-momentum tensor and  
a contribution from its current density. 

The potential $\psi^A$ is singular on the world line, and so is
$\nabla_\mu \psi^A$, which enters the equations of motion satisfied by
the particle. These singularities are now well understood and easily
tamed, and because Eq.~(\ref{eq1}) has the same mathematical structure
as the perturbation equations examined in the past, they can be
handled with the same regularization techniques. Here we follow the
approach advocated by Detweiler and Whiting
\cite{detweiler-whiting:03}: we decompose the retarded 
potential $\psi^A$ into a precisely-defined singular potential 
$\psi^A_{\sf S}$ and a regular remainder $\psi^A_{\sf R}$, and assert
that only $\psi^A_{\sf R}$ appears in the equations of motion. The
manipulations associated with the decomposition can all be performed
at the level of Eqs.~(\ref{eq1}) and (\ref{eq2}), and the behavior of
the potentials near the world line can also be obtained directly from
these equations. For these purposes there is no need to specify the
precise identity of the potentials $\psi^A$, and there is no need to
identify the tensors $M^A_{\ B\mu}$ and $N^A_{\ B}$; the only
important aspect is that $\psi^A$ is a solution to Eq.~(\ref{eq1}).   

The methods developed here to handle coupled perturbations in
nonvacuum spacetimes are therefore quite general, and they apply well
beyond the specific context of gravitational and electromagnetic
perturbations of an electrovac background spacetime. We do apply the
formalism to this specific situation, but we also consider the case of a
particle with scalar charge $q$ moving in a background spacetime that
contains a background scalar field. (We name ``scalarvac spacetime''
such a solution to the Einstein-scalar field equations.) Other
applications, for example, a computation of the gravitational
self-force in a scalar-tensor theory of gravity, could also benefit
from the techniques developed here.  

We begin in Sec.~\ref{sec:scalarvac} with a derivation of the
perturbation equations in the case of a scalarvac spacetime perturbed
by a particle of mass $m$ and scalar charge $q$. In
Sec.~\ref{sec:electrovac} we turn to the perturbation equations for an  
electrovac spacetime perturbed by a particle of mass $m$ and electric
charge $e$. In Sec.~\ref{sec:meta} we cast the perturbation equations
in the general form of Eq.~(\ref{eq1}) and provide additional details
regarding the meta-index notation. In Sec.~\ref{sec:Hadamard} we
introduce the Green's function $G^A_{\ B'}(x,x')$, and construct its
local Hadamard expansion when $x$ is sufficiently close to $x'$. In
Sec.~\ref{sec:potentials} we examine the retarded, singular, and
regular potentials near the world line, and collect the essential
ingredients required in the computation of the self-force. The
self-force acting on a particle of mass $m$ and scalar charge $q$
moving in a background scalarvac spacetime is calculated in
Sec.~\ref{sec:scalar_sf}; our final result is displayed in
Eqs.~(\ref{eom_scalar5}) and (\ref{eom_scalar6}), and expressed in
terms of ``tail fields'' defined in Eqs.~(\ref{scalar_phi_tail}),
(\ref{scalar_h_tail}), and (\ref{scalar_grad_tail}). The self-force
acting on particle of mass $m$ and electric charge $e$ moving in an
electrovac spacetime is calculated in Sec.~\ref{sec:em_sf}; our final
result is displayed in Eq.~(\ref{eom_em3}), and the tail fields are
defined in Eqs.~(\ref{em_b_tail}), (\ref{em_h_tail}), and
(\ref{em_grad_tail}). 

In both the scalarvac and electrovac cases the equations of motion
take the schematic form of 
\begin{equation} 
m a^\mu = F^\mu_{\rm back}
+ F^\mu_{\rm local} + F^\mu_{\rm tail}, 
\label{eq3} 
\end{equation} 
where $a^\mu$ is the covariant acceleration, $F^\mu_{\rm back}$
is the force exerted by the background field in the background
spacetime, $F^\mu_{\rm local}$ is the local piece of the self-force,
which depends on tensors defined in the background spacetime and
evaluated at the current position of the particle, and 
$F^\mu_{\rm tail}$ is the tail piece of the self-force, which depends
on the past history of the particle. Because a formal expression in
terms of local and tail pieces is rarely a useful starting point for
the explicit evaluation of the self-force, we also provide covariant
expressions for the singular potential $\psi^A_{\sf S}$ expressed as a
local expansion near the world line. These expressions can be found in 
Secs.~\ref{subsec:sing_general}, \ref{subsec:sing_scalar}, and
\ref{subsec:sing_em}. The singular potential can be involved in the
construction of an effective extended source for the regular
potential \cite{vega-wardell-diener:11, diener-etal:12,
  wardell-etal:12}, or in the computation of regularization
parameters \cite{barack-ori:02, barack-ori:03a, barack-etal:02}  
when the self-force is computed as an infinite sum over
spherical-harmonic modes. We do not pursue such computations 
here, but refer the reader to an independent effort\footnote{Thomas
  M. Linz, John L. Friedman, Alan G. Wiseman, in preparation.} to
calculate the singular part of the self-force acting on a charged
particle moving in an electrovac spacetime.   

As mentioned previously, our derivation of the self-force in
nonvacuum spacetimes rests on the Detweiler-Whiting prescription to
regularize the potentials of a point particle, which diverge on the
world line. Another set of methods, based on effective field theory
\cite{galley-hu-lin:06, galley-hu:09}, can also be adopted, and this
shall be explored in a forthcoming publication.\footnote{Peter
  Zimmerman, in preparation.}   
 
Our developments in this paper rely heavily on the general theory of
bitensors in curved spacetime, the theory of Green's functions and
their Hadamard representation, the description of a neighborhood of a
world line in terms of Fermi normal coordinates, and a host of other
techniques that have become standard fare in the self-force
literature. For all this we refer the reader to the comprehensive
review of Poisson, Pound, and Vega \cite{poisson-pound-vega:11}, to
which we repeatedly refer as PPV.  

\section{Perturbed scalarvac spacetimes} 
\label{sec:scalarvac} 

We consider a background spacetime whose metric $g_{\alpha\beta}$ is a 
solution to the Einstein field equations in the presence of a scalar
field $\Phi$. With a suitable normalization for the scalar field, the
action functional for the system is 
\begin{equation} 
S = \frac{1}{16\pi} \int R\, dV 
- \frac{1}{8\pi} \int \biggl( \frac{1}{2} g^{\alpha\beta}
\nabla_\alpha \Phi \nabla_\beta \Phi + F \biggr)\, dV,
\end{equation} 
where $R$ is the Ricci scalar, $dV := \sqrt{-g}\, d^4x$ is the
invariant volume element, and $F(\Phi)$ is a potential for the scalar
field. Variation of the action with respect to the metric yields the
field equations 
\begin{equation} 
G_{\alpha\beta} = 8\pi T_{\alpha\beta} 
= \nabla_\alpha \Phi \nabla_\beta \Phi 
- \biggl( \frac{1}{2} \nabla^\mu \Phi \nabla_\mu \Phi + F \biggr)
g_{\alpha\beta}. 
\label{grav_scalar_FE} 
\end{equation} 
They imply  
\begin{equation} 
R_{\alpha\beta} = \nabla_\alpha \Phi \nabla_\beta \Phi 
+ F g_{\alpha\beta}, \qquad 
R = \nabla^\alpha \Phi \nabla_\alpha \Phi + 4 F. 
\label{Ricci_scalar} 
\end{equation} 
Variation of the action with respect to the scalar field produces the
wave equation  
\begin{equation} 
\Box \Phi = F' 
\label{scalar_FE} 
\end{equation} 
for the scalar field, where 
$\Box := g^{\alpha\beta} \nabla_\alpha \nabla_\beta$ is the covariant
wave operator and $F' := dF/d\Phi$.  

The background spacetime is perturbed by a point particle carrying a
scalar charge $q$ and a mass $m$; the particle moves on a word line
$\gamma$ described by the parametric relations $z^\mu(\tau)$, in
which $\tau$ is proper time in the background spacetime. This is
achieved by adding 
\begin{equation} 
S_{\rm pert} = -m \int_\gamma\, d\tau 
+ q \int_\gamma \Phi\bigl(z(\tau)\bigr)\, d\tau 
= -\int_\gamma (m-q\Phi)\, d\tau 
\end{equation} 
to the background action; the first term is the action of a free
particle in a curved spacetime, and the second term accounts for its
interaction with the scalar field. The perturbed
metric is ${\sf g}_{\alpha\beta} = g_{\alpha\beta} + h_{\alpha\beta}$,
where $h_{\alpha\beta}$ is the metric perturbation, and the perturbed
scalar field is ${\sf \Phi} = \Phi + \phi$, where $\phi$ is the
perturbation. We introduce 
\begin{equation} 
\gamma_{\alpha\beta} := h_{\alpha\beta} - \frac{1}{2} g_{\alpha\beta} h 
\label{gamma_def} 
\end{equation}  
as a ``trace-reversed'' perturbation, where $h :=
g^{\alpha\beta} h_{\alpha\beta}$. It is understood that indices
on $h_{\alpha\beta}$ and $\gamma_{\alpha\beta}$ are raised with the
inverse background metric $g^{\alpha\beta}$. 

The perturbation is sourced by the particle's energy-momentum tensor
$t^{\alpha\beta}$ and its scalar-charge density $j$. The
energy-momentum tensor is obtained by varying $S_{\rm pert}$ with
respect to the metric, and we get  
\begin{equation} 
t^{\alpha\beta}(x) = \int_\gamma (m - q\Phi) g^{\alpha}_{\ \mu}(x,z) 
g^{\beta}_{\ \nu}(x,z) u^\mu u^\nu \delta_4(x,z)\, d\tau,  
\label{t_def}
\end{equation} 
where $\Phi := \Phi(z)$, $u^\mu := dz^\mu/d\tau$ is the particle's
velocity vector, $g^\alpha_{\ \mu}(x,z)$ is the parallel propagator from
$z(\tau)$ to $x$, and $\delta_4(x,z)$ is a scalarized Dirac
distribution. The scalar-charge density is obtained by varying 
$S_{\rm pert}$ with respect to $\Phi$, and we get
\begin{equation} 
j(x) = q \int_\gamma  \delta_4(x,z)\, d\tau.  
\label{j_def} 
\end{equation} 

We denote ${\sf G}_{\alpha\beta}$ the Einstein tensor of the perturbed
spacetime. Its perturbation $\delta G_{\alpha\beta} 
= {\sf G}_{\alpha\beta} - G_{\alpha\beta}$ relative to the background
Einstein tensor is given by  
\begin{equation} 
\delta G_{\alpha\beta} = \frac{1}{2} \biggl( 
-\Box \gamma_{\alpha\beta} 
+ \nabla_\alpha \nabla_\mu \gamma^\mu_{\ \beta} 
+ \nabla_\beta \nabla_\mu \gamma^\mu_{\ \alpha} 
- g_{\alpha\beta} \nabla_\mu \nabla_\nu \gamma^{\mu\nu}  
- 2 R^{\mu\ \nu}_{\ \alpha\ \beta} \gamma_{\mu\nu} 
+ R_{\alpha\mu} \gamma^\mu_{\ \beta} 
+ R_{\beta\mu} \gamma^\mu_{\ \alpha}
+ g_{\alpha\beta} R_{\mu\nu} \gamma^{\mu\nu} 
- R \gamma_{\alpha\beta} \biggr), 
\label{dG} 
\end{equation} 
in which $\nabla_{\alpha}$ indicates covariant differentiation in the
background spacetime. Similarly, we denote ${\sf T}_{\alpha\beta}$ the
energy-momentum tensor of the perturbed scalar field in the perturbed
spacetime. Its perturbation $\delta T_{\alpha\beta} 
= {\sf T}_{\alpha\beta} - T_{\alpha\beta}$ is given by  
\begin{equation} 
8\pi \delta T_{\alpha\beta} = \nabla_\alpha \Phi \nabla_\beta \phi
+ \nabla_\beta \Phi \nabla_\alpha \phi
 - g_{\alpha\beta} \nabla^\mu \Phi \nabla_\mu \phi 
- g_{\alpha\beta} F' \phi 
+ \frac{1}{2} g_{\alpha\beta} \nabla_\mu \Phi \nabla_\nu \Phi
\gamma^{\mu\nu} 
- \biggl( \frac{1}{2} \nabla^\mu \Phi \nabla_\mu \Phi 
+ F \biggr) \gamma_{\alpha\beta} 
+ \frac{1}{2} g_{\alpha\beta} F \gamma, 
\label{dT_scalar} 
\end{equation} 
in which $F$ and $F'$ are evaluated at the background field $\Phi$. 

The Einstein field equations ${\sf G}_{\alpha\beta} = 
8\pi ({\sf T}_{\alpha\beta}  + t_{\alpha\beta})$ become 
$\delta G_{\alpha\beta} - 8\pi \delta T_{\alpha\beta} 
= 8\pi t_{\alpha\beta}$ after linearization. The equations simplify
(and become manifestly hyperbolic) when the gravitational perturbation
is required to satisfy the one-parameter family of gauge conditions 
\begin{equation} 
\nabla_\beta \gamma^{\alpha\beta} = 2\lambda \phi \nabla_\alpha \Phi, 
\label{lambda-gauge} 
\end{equation} 
in which $\lambda$ is a free (dimensionless) parameter. Making the 
substitutions from Eqs.~(\ref{dG}) and (\ref{dT_scalar}) and
exploiting Eqs.~(\ref{Ricci_scalar}) returns the wave equation 
\begin{equation} 
\Box \gamma^{\alpha\beta} 
+ M^{\alpha\beta}_{\ \ \ |\cdot\mu} \nabla^\mu \phi 
+ N^{\alpha\beta}_{\ \ \ |\gamma\delta} \gamma^{\gamma\delta} 
+ N^{\alpha\beta}_{\ \ \ |\cdot} \phi 
= -16\pi t^{\alpha\beta} 
\label{wave_grav_scalar} 
\end{equation} 
for the gravitational perturbation. We have introduced the tensors 
\begin{subequations} 
\label{MN-scalar1} 
\begin{align} 
M^{\alpha\beta}_{\ \ \ |\cdot\mu}&:= 2(1-\lambda) \bigl( 
\delta^{\alpha}_{\ \mu} \nabla^\beta \Phi 
+ \delta^{\beta}_{\ \mu} \nabla^\alpha \Phi 
- g^{\alpha\beta} \nabla_\mu \Phi \bigr), \\ 
N^{\alpha\beta}_{\ \ \ |\gamma\delta}&= 
2 R^{\alpha\ \, \beta}_{\ (\gamma\ \delta)} 
- \delta^\alpha_{\ (\gamma} \nabla^\beta \Phi \nabla_{\delta)} \Phi 
- \delta^\beta_{\ (\gamma} \nabla^\alpha \Phi \nabla_{\delta)} \Phi, \\ 
N^{\alpha\beta}_{\ \ \ |\cdot} &= -2 \bigl[ 
(1-\lambda) g^{\alpha\beta} F'  
+ 2\lambda \nabla^{\alpha} \nabla^\beta \Phi \bigr].  
\end{align} 
\end{subequations} 
The notation for the $M$ and $N$ tensors, involving vertical stokes
and dots, will be explained fully below. Notice that 
$M^{\alpha\beta}_{\ \ \ |\mu} = 0$ when $\lambda = 1$;  in this gauge
the wave equation does not involve first-derivative terms in $\phi$. 

The scalar field equation ${\sf g}^{\alpha\beta} {\sf D}_\alpha
{\sf D}_{\beta} {\sf \Phi} = F'({\sf \Phi}) - 4\pi j$, in which 
${\sf D}_{\alpha}$ indicates covariant differentiation in the
perturbed spacetime, becomes 
\begin{equation} 
\Box \phi + N^{\cdot}_{\ |\alpha\beta} \gamma^{\alpha\beta} 
+ N^{\cdot}_{\ |\cdot} \phi = -4\pi j 
\label{wave_scalar} 
\end{equation} 
after linearization. Here 
\begin{subequations} 
\label{MN-scalar2} 
\begin{align} 
N^{\cdot}_{\ |\alpha\beta} &= -\biggl( \nabla_\alpha \nabla_\beta \Phi 
- \frac{1}{2} F' g_{\alpha\beta} \biggr), \\  
N^{\cdot}_{\ |\cdot} &= -\bigl( 2\lambda \nabla^\gamma \Phi
\nabla_\gamma \Phi + F'' \bigr). 
\end{align}
\end{subequations} 
This equation does not feature a coupling to $\nabla^\mu
\gamma^{\alpha\beta}$, irrespective of the choice of gauge. 

\section{Perturbed electrovac spacetimes} 
\label{sec:electrovac} 

In this section the background spacetime contains an electromagnetic
field $F_{\alpha\beta}$ instead of a scalar field $\Phi$. The action
functional is now 
\begin{equation} 
S = \frac{1}{16\pi} \int R\, dV 
- \frac{1}{16\pi} \int F_{\alpha\beta} F^{\alpha\beta}\, dV, 
\end{equation} 
and the Einstein field equations are 
\begin{equation} 
G_{\alpha\beta} = 8\pi T_{\alpha\beta} 
= 2 F_\alpha^{\ \gamma} F_{\beta\gamma} 
- \frac{1}{2} g_{\alpha\beta} F^{\gamma\delta} F_{\gamma\delta};  
\label{grav_em_FE} 
\end{equation} 
they imply $R_{\alpha\beta} = G_{\alpha\beta}$, $R = 0$. Maxwell's
equations are  
\begin{equation} 
\nabla_\beta F^{\alpha\beta} = 0, \qquad 
\nabla_\alpha F_{\beta\gamma} 
+ \nabla_\gamma F_{\alpha\beta} 
+ \nabla_\beta F_{\gamma\alpha} = 0; 
\label{em_FE} 
\end{equation} 
we specifically assume that there is no source for the background
electromagnetic field.  

The background spacetime is perturbed by a point particle carrying an
electric charge $e$ and a mass $m$; the particle moves on a word line
$\gamma$ described by the parametric relations $z^\mu(\tau)$, in
which $\tau$ is proper time in the background spacetime. This is
achieved by adding 
\begin{equation} 
S_{\rm pert} = -m \int_\gamma\, d\tau 
+ e \int_\gamma A_\mu u^\mu\, d\tau  
\end{equation} 
to the action functional, where $A_\mu$ is a vector potential for the
electromagnetic field. The metric perturbation was introduced
previously, and the perturbed electromagnetic field is 
${\sf F}_{\alpha\beta} = F_{\alpha\beta}  + f_{\alpha\beta}$, with
$f_{\alpha\beta}$ denoting the perturbation. It is useful to introduce
a second vector potential $b_\alpha$ such that 
\begin{equation} 
f_{\alpha\beta} = \nabla_\alpha b_\beta - \nabla_\beta b_\alpha. 
\label{b_def}
\end{equation} 
It is understood that indices on $b_{\alpha}$ are raised with the
inverse background metric $g^{\alpha\beta}$. 

The perturbation is sourced by the particle's energy-momentum tensor
$t^{\alpha\beta}$, which is now given by 
\begin{equation} 
t^{\alpha\beta}(x) = m \int_\gamma g^{\alpha}_{\ \mu}(x,z) 
g^{\beta}_{\ \nu}(x,z) u^\mu u^\nu \delta_4(x,z)\, d\tau,  
\end{equation} 
and by the current density
\begin{equation} 
j^\alpha(x) = e \int_\gamma  g^\alpha_{\ \mu}(x,z) u^\mu 
\delta_4(x,z)\, d\tau.  
\label{jem_def} 
\end{equation} 
The perturbation of the Einstein tensor is still given by
Eq.~(\ref{dG}), and the perturbation of the electromagnetic
energy-momentum tensor is 
\begin{equation} 
4\pi \delta T_{\alpha\beta} = F_{\alpha}^{\ \mu} f_{\beta\mu} 
+ F_{\beta}^{\ \mu} f_{\alpha\mu}  
- \frac{1}{2} g_{\alpha\beta} F^{\mu\nu} f_{\mu\nu} 
- F_{\alpha\mu} F_{\beta\nu} \gamma^{\mu\nu} 
+ \frac{1}{2} g_{\alpha\beta} F_{\mu}^{\ \lambda}
F_{\nu\lambda} \gamma^{\mu\nu} 
- \frac{1}{4} F^{\mu\nu} F_{\mu\nu} \gamma_{\alpha\beta} 
+ (2\pi T_{\alpha\beta}) \gamma. 
\end{equation} 

As in the previous section we have that the linearized Einstein field 
equations are $\delta G_{\alpha\beta} - 8\pi \delta T_{\alpha\beta} =
8\pi t_{\alpha\beta}$, and these lead to a set of differential
equations for the metric perturbation $\gamma_{\alpha\beta}$. 
Maxwell's equations ${\sf D}^\beta {\sf F}_{\alpha\beta} 
= 4\pi j_\alpha$ and ${\sf D}_\alpha {\sf F}_{\beta\gamma} 
+ {\sf D}_\gamma {\sf F}_{\alpha\beta} + 
{\sf D}_\beta {\sf F}_{\gamma\alpha} = 0$ can similarly be cast in
the form of differential equations for $b^\alpha$. To
simplify the linearized field equations we impose a Lorenz-gauge 
condition on both $b_\alpha$ and $\gamma_{\alpha\beta}$: 
\begin{equation} 
\nabla_\alpha b^\alpha = 0, \qquad 
\nabla_\beta \gamma^{\alpha\beta} = 0. 
\label{lorenz} 
\end{equation} 
After some manipulations we find that the perturbation
equations become  
\begin{equation} 
\Box \gamma^{\alpha\beta} 
+ M^{\alpha\beta}_{\ \ \ |\gamma\mu}  \nabla^\mu b^\gamma 
+ N^{\alpha\beta}_{\ \ \ |\gamma\delta} \gamma^{\gamma\delta} 
= -16 \pi t^{\alpha\beta} 
\label{wave_grav_em} 
\end{equation} 
and 
\begin{equation} 
\Box b^\alpha 
+ M^\alpha_{\ \ |\beta\gamma\mu} \nabla^\mu \gamma^{\beta\gamma}  
+ N^\alpha_{\ \ |\beta\gamma} \gamma^{\beta\gamma} 
+ N^\alpha_{\ \ |\beta} b^\beta 
= -4\pi j^\alpha,  
\label{wave_em} 
\end{equation} 
with 
\begin{subequations} 
\label{MN_em1} 
\begin{align} 
M^{\alpha\beta}_{\ \ \ |\gamma\mu} &= 
16 \delta^{(\alpha}_{\ [\mu} F^{\beta)}_{\ \gamma]} 
- 4 g^{\alpha\beta} F_{\mu\gamma}, \\ 
N^{\alpha\beta}_{\ \ \ |\gamma\delta} &= 2 \bigl( 
R^{\alpha\ \, \beta}_{\ (\gamma\ \delta)} 
- \delta^\alpha_{\ (\gamma} F^{\beta\epsilon} F_{\delta)\epsilon} 
- \delta^\beta_{\ (\gamma} F^{\alpha\epsilon} F_{\delta)\epsilon} 
- 2 F^\alpha_{\ (\gamma} F^\beta_{\ \delta)} 
+ g_{\gamma\delta} F^{\alpha\epsilon} F^\beta_{\ \epsilon} \bigr)
\end{align} 
\end{subequations} 
and 
\begin{subequations} 
\label{MN_em2} 
\begin{align} 
M^\alpha_{\ \ |\beta\gamma\mu} &= 
\delta^\alpha_{\ (\beta} F_{\gamma)\mu}  
- \frac{1}{2} F^\alpha_{\ \mu} g_{\beta\gamma}, \\ 
N^\alpha_{\ \ |\beta\gamma} &= \nabla_{(\beta} F^\alpha_{\ \gamma)}\,
\\  
N^\alpha_{\ \ |\beta} &= -R^\alpha_{\ \beta}. 
\end{align} 
\end{subequations} 
There is no choice of gauge that permits the elimination of 
$M^{\alpha\beta}_{\ \ \ |\beta\gamma\mu}$ and  
$M^\alpha_{\ \ |\beta\gamma\mu}$ from the perturbation equations;
these necessarily contain first-derivative terms in both 
$\gamma_{\alpha\beta}$ and $b_\alpha$.  

\section{Condensed index notation} 
\label{sec:meta} 

To integrate the linearized field equations displayed in the preceding
sections, it is convenient to handle them all at once by exploiting a
condensed index notation for the various fields. We thus introduce a
meta-index $A$ which can stand for three types of tensorial
indices. First, $A$ can stand for a pair of indices,  
$A = \alpha\beta$,  which is understood to be symmetrized. Second,
$A$ can stand for a single index, $A = \alpha$. And third, $A$ can
stand for an absence of index, $A = \cdot$. In this way we can
collect the tensor field $\gamma^{\alpha\beta}$, the vector field
$b^\alpha$, and the scalar field $\phi$ into the single meta-object 
$\psi^A$. 

In a similar way we collect the source terms $4 t^{\alpha\beta}$,
$j^\alpha$, and $j$ into the meta-object $\mu^A$. The perturbation  
equations can then be expressed as 
\begin{equation} 
\Box \psi^A + M^A_{\ B\mu} \nabla^\mu \psi^B 
+ N^A_{\ B} \psi^B = -4\pi \mu^A, 
\label{metaFE} 
\end{equation} 
where summation over repeated meta-indices is understood. Such sums
include all possible combinations of indices; for example, when  
Eq.~(\ref{metaFE}) represents Eq.~(\ref{wave_em}) and 
$A$ stands for $\alpha$, the summation over $B$ in 
$N^A_{\ B} \psi^B$ stands explicitly for 
$N^\alpha_{\ \ |\beta\gamma} \gamma^{\beta\gamma} 
+ N^\alpha_{\ \ |\beta} b^\beta$. The various tensor fields 
$M^A_{\ B\mu}$ and  $N^A_{\ B}$ can be read off from
Eqs.~(\ref{MN-scalar1}), (\ref{MN-scalar2}), (\ref{MN_em1}), and
(\ref{MN_em2}). In the more explicit notation used 
in these equations, a vertical stroke separates the indices collected
in $A$ from those collected in $B$; for example, $N^A_{\ B}$ is denoted
$N^{\alpha\beta}_{\ \ \ |\cdot}$ when $A$ stands for $\alpha\beta$ and
$B$ stands for $\cdot$. 

We next introduce other useful meta-objects. We set 
\begin{equation} 
q^A := \left\{ 
\begin{array}{l} 
4 (m-q\Phi) u^\alpha u^\beta \\  
e u^\alpha \\ 
q  
\end{array}  
\right. ,
\label{q_def} 
\end{equation} 
\begin{equation} 
\dot{q}^A := \left\{ 
\begin{array}{l} 
8 (m-q\Phi) a^{(\alpha} u^{\beta)} 
- 4q \dot{\Phi} u^\alpha u^\beta \\  
e a^\alpha \\ 
0
\end{array}  
\right. ,
\label{qdot_def} 
\end{equation} 
\begin{equation} 
\ddot{q}^A := \left\{ 
\begin{array}{l} 
8 (m-q\Phi) \bigl( \dot{a}^{(\alpha} u^{\beta)} 
+ a^\alpha a^\beta \bigr) 
- 16 q \dot{\Phi} a^{(\alpha} u^{\beta)} 
- 4 q \ddot{\Phi} u^\alpha u^\beta \\  
e \dot{a}^\alpha \\ 
0
\end{array}  
\right. ,
\label{qddot_def} 
\end{equation} 
\begin{equation} 
\delta^A_{\ B} := \left\{ 
\begin{array}{l} 
\delta^{(\alpha}_{\ \gamma} \delta^{\beta)}_{\ \delta} \\ 
\delta^\alpha_{\ \beta} \\ 
1 
\end{array} 
\right. ,
\end{equation} 
and 
\begin{equation} 
g^A_{\ B'}(x,x') := \left\{ 
\begin{array}{l} 
g^{(\alpha}_{\ \gamma'}(x,x') g^{\beta)}_{\ \delta'}(x,x') \\ 
g^\alpha_{\ \beta'}(x,x') \\ 
1
\end{array} 
\right. .
\label{gAB_def} 
\end{equation} 
With $\delta^A_{\ B}$ and $g^A_{\ B'}$ it is understood that the 
object is zero whenever $A$ and $B$ (or $B'$) stand for different
types of indices; for example $\delta^{\alpha}_{\ |\cdot} = 0$ and 
$g^{\alpha\beta}_{\ \ \ |\gamma'} = 0$. We have introduced $a^\alpha
:= D u^\alpha/d\tau$ as the particle's acceleration vector, and
$\dot{a}^\alpha := D a^\alpha/d\tau$ as its rate of change along 
the world line. We also denote $\dot{\Phi} := u^\mu \nabla_\mu \Phi$, 
$\ddot{\Phi} := u^\nu \nabla_\nu (u^\mu \nabla_\mu \Phi)$.   

With these objects we find that Eqs.~(\ref{t_def}), (\ref{j_def}), and
(\ref{jem_def}) are all contained in the meta-expression 
\begin{equation} 
\mu^A(x) = \int_\gamma g^A_{\ M}(x,z) q^M(\tau) \delta_4(x,z)\, d\tau. 
\label{mu_def} 
\end{equation} 
Another meta-object we shall need in further developments is 
\begin{equation} 
R^A_{\ B'\mu'\nu'} := \left\{ 
\begin{array}{l} 
g^{(\alpha}_{\ \alpha'} g^{\beta)}_{\ \gamma'} 
R^{\alpha'}_{\ \delta'\mu'\nu'} 
+ g^{(\alpha}_{\ \alpha'} g^{\beta)}_{\ \delta'} 
R^{\alpha'}_{\ \gamma'\mu'\nu'} \\ 
g^{\alpha'}_{\ \alpha'} R^{\alpha'}_{\ \beta'\mu'\nu'} \\ 
0
\end{array} 
\right. ,
\label{RAB_def} 
\end{equation} 
in which the Riemann tensor is evaluated at $x'$, and the parallel
propagators refer to both $x$ and $x'$. Here also it is understood
that $R^A_{\ B'\mu'\nu'}$ is zero when $A$ and $B'$ stand for
different types of indices. 

\section{Green's function and Hadamard expansion} 
\label{sec:Hadamard} 

To integrate Eq.~(\ref{metaFE}) we introduce a Green's function
$G^A_{\ B'}(x,x')$ that satisfies 
\begin{equation} 
\Box G^A_{\ B'} + M^A_{\ B\mu} \nabla^\mu G^B_{\ B'} 
+ N^A_{\ B} G^B_{\ B'} = -4\pi g^A_{\ B'} \delta_4(x,x') 
\label{metaGreen} 
\end{equation} 
together with suitable boundary conditions; we shall be exclusively
concerned with the retarded Green's function, which vanishes when 
$x$ is in the past of $x'$. The solution to the wave equation can then
be expressed as  
\begin{equation} 
\psi^A(x) = \int G^A_{\ B'}(x,x') \mu^{B'}(x')\, dV', 
\label{psi_sol1} 
\end{equation} 
where $dV' := \sqrt{-g'}\, d^4x'$ is the element of spacetime volume
at $x'$. With the source term of Eq.~(\ref{mu_def}) this becomes 
\begin{equation} 
\psi^A(x) = \int_\gamma G^A_{\ M}(x,z) q^M(\tau)\, d\tau,
\label{psi_sol2} 
\end{equation} 
in which the Green's function is evaluated at $x'=z(\tau)$. It must be
remembered that this equation contains a summation over the 
repeated meta-index $M$. For example, when Eq.~(\ref{psi_sol2})
represents the solution to the gravity-scalar wave equation
(\ref{wave_grav_scalar}), its explicit expression is 
\begin{equation} 
\gamma^{\alpha\beta}(x) = 4\int_\gamma (m-q\Phi) 
G^{\alpha\beta}_{\ \ \ |\mu\nu}(x,z) u^{\mu} u^{\nu}\, d\tau 
+ q \int_\gamma G^{\alpha\beta}_{\ \ \ |\cdot}(x,z)\, d\tau. 
\end{equation} 

When $x$ is within the normal convex neighborhood of $x'$, the
retarded Green's function can be cast in the Hadamard form 
\begin{equation} 
G^A_{\ B'}(x,x') = U^A_{\ B'}(x,x') \delta_+(\sigma) 
+ V^A_{\ B'}(x,x') \Theta_+(-\sigma), 
\label{Hadamard} 
\end{equation} 
in which $\sigma(x,x')$ is Synge's world function, $\delta_+(\sigma)$ 
and $\Theta_+(-\sigma)$ are respectively the Dirac and Heaviside
distributions restricted to the future of $x'$, and $U^A_{\ B'}(x,x')$
and $V^A_{\ B'}(x,x')$ are smooth bitensors. Following the steps
detailed in Sec.~14.2 of PPV \cite{poisson-pound-vega:11}, we find
that $U^A_{\ B'}$ satisfies the differential equation  
\begin{equation} 
2 \sigma^\mu \nabla_\mu U^A_{\ B'} 
+ \sigma^\mu M^A_{\ B\mu} U^B_{\ B'} 
+ \bigl( \sigma^\mu_{\ \mu} - 4 \bigr) U^A_{\ B'} = 0 
\label{U_deq} 
\end{equation} 
together with the boundary conditions 
\begin{equation} 
U^A_{\ B'}(x',x') = g^A_{\ B'}(x',x') = \delta^{A'}_{\ B'}. 
\label{U_bc} 
\end{equation} 
We also find that $V^A_{\ B'}$ satisfies 
\begin{equation} 
2 \sigma^\mu \nabla_\mu V^A_{\ B'} 
+ \sigma^\mu M^A_{\ B\mu} V^B_{\ B'} 
+ \bigl( \sigma^\mu_{\ \mu} - 2 \bigr) V^A_{\ B'} = 
\Box U^A_{\ B'} + M^A_{\ B\mu} \nabla^\mu U^B_{\ B'} 
+ N^A_{\ B} U^B_{\ B'} 
\label{V_deq1} 
\end{equation} 
on the light cone $\sigma(x,x') = 0$, as well as the wave equation  
\begin{equation} 
\Box V^A_{\ B'} + M^A_{\ B\mu} \nabla^\mu V^B_{\ B'} 
+ N^A_{\ B} V^B_{\ B'} = 0 
\label{V_deq2} 
\end{equation}
everywhere. In these equations, $\sigma_\mu$ stands for a partial
derivative of $\sigma(x,x')$ with respect to $x^\mu$, and
$\sigma_{\mu\nu} := \nabla_\mu \nabla_\nu \sigma(x,x')$ --- here also
the derivatives are taken with respect to $x$. 

Equations (\ref{U_deq}) and (\ref{U_bc}) allow us to construct 
$U^A_{\ B}(x,x')$ as a covariant expansion in powers of
$\sigma^{\mu'}$ about $x'$. We write 
\begin{equation} 
U^A_{\ B'}(x,x') = g^A_{\ A'} \biggl[ \delta^{A'}_{\ B'} 
+ U^{A'}_{\ B'\mu'} \sigma^{\mu'} 
+ \frac{1}{2} U^{A'}_{\ B'\mu'\nu'} \sigma^{\mu'} \sigma^{\nu'} 
+ O(\epsilon^3) \biggr], 
\label{U_taylor1}
\end{equation} 
in which $U^{A'}_{\ B'\mu'}$ and $U^{A'}_{\ B'\mu'\nu'}$ are ordinary
tensors at $x'$, and $\epsilon$ is a measure of distance between
$x$ and $x'$. By inserting this expression in Eq.~(\ref{U_deq}) and
solving order-by-order in $\sigma^{\mu'}$, we arrive at 
\begin{subequations} 
\label{U_taylor2} 
\begin{align} 
U^{A'}_{\ B'\mu'} &= \frac{1}{2} M^{A'}_{\ B'\mu'}, \\ 
U^{A'}_{\ B'\mu'\nu'} &= 
-\frac{1}{2} \nabla_{(\mu'} M^{A'}_{\ B'\nu')} 
+ \frac{1}{4} M^{A'}_{\ C'(\mu'} M^{C'}_{\ B'\nu')} 
+ \frac{1}{6} \delta^{A'}_{\ B'} R_{\mu'\nu'},  
\end{align} 
\end{subequations} 
where $M^{A'}_{\ B'\mu'}$ is the tensor field $M^A_{\ B\mu}$
evaluated at $x'$. In the expression for $U^{A'}_{\ B'\mu'\nu'}$ the
indices contained in $B'$ are excluded from the symmetrization over
the $\mu'$ and $\nu'$ indices, and summation over $C'$ involves all
possible combinations of indices. The manipulations leading to 
Eqs.~(\ref{U_taylor2}) rely on standard identities among bitensors,
such as $\sigma_{\mu'} = -g^{\mu}_{\ \mu'} \sigma_{\mu}$, 
$\sigma^\mu \nabla_\mu g^\alpha_{\ \alpha'} = 0$ (PPV Sec.~5.4),  
and the expansion $\sigma^\mu_{\ \mu} = 4 
- \frac{1}{3} R_{\mu'\nu'} \sigma^{\mu'} \sigma^{\nu'} 
+ O(\epsilon^3)$ (PPV Sec.~6.2).

From Eq.~(\ref{U_taylor1}) we may compute derivatives of 
$U^A_{\ B'}$. For this purpose we use the expansions (PPV Sec.~6.2)  
\begin{equation} 
\nabla_{\mu'} g^{\alpha}_{\ \beta'} = \frac{1}{2} g^\alpha_{\ \alpha'} 
R^{\alpha'}_{\ \beta'\mu'\nu'} \sigma^{\nu'} + O(\epsilon^2), \qquad 
\nabla_{\mu} g^{\alpha}_{\ \beta'} = \frac{1}{2} g^\alpha_{\ \alpha'} 
g^{\mu'}_{\ \mu} R^{\alpha'}_{\ \beta'\mu'\nu'} \sigma^{\nu'} 
+ O(\epsilon^2)  
\end{equation} 
to show that with $g^A_{\ B'}$ defined by Eq.~(\ref{gAB_def}) and
$R^A_{\ B'\mu'\nu'}$ defined by Eq.~(\ref{RAB_def}), 
\begin{equation} 
\nabla_{\mu'} g^A_{\ B'} = \frac{1}{2} R^A_{\ B'\mu'\nu'}
\sigma^{\nu'} + O(\epsilon^2) , \qquad 
\nabla_{\mu} g^A_{\ B'} = \frac{1}{2} g^{\mu'}_{\ \mu} 
R^A_{\ B'\mu'\nu'} \sigma^{\nu'} + O(\epsilon^2). 
\label{grad_gAB} 
\end{equation} 
With this and the standard expansions for $\sigma^{\nu'}_{\ \mu'}$ and
$\sigma^{\nu'}_{\ \mu}$ displayed in Sec.~6.2 of PPV, we obtain 
\begin{equation} 
\nabla_{\mu'} U^A_{\ B'} = g^A_{\ A'} \Bigl[ U^{A'}_{\ B'\mu'} 
+ \bigl( \nabla_{\mu'} U^{A'}_{\ B'\nu'} + U^{A'}_{\ B'\mu'\nu'}
\bigr) \sigma^{\nu'} \Bigr] 
+ \frac{1}{2} R^A_{\ B'\mu'\nu'} \sigma^{\nu'} + O(\epsilon^2)  
\label{gradUAB1} 
\end{equation} 
and 
\begin{equation} 
\nabla_{\mu} U^A_{\ B'} = - g^{\mu'}_{\ \mu} g^A_{\ A'} 
\Bigl[ U^{A'}_{\ B'\mu'} 
+ U^{A'}_{\ B'\mu'\nu'} \sigma^{\nu'} \Bigr] 
+ \frac{1}{2} g^{\mu'}_{\ \mu} R^A_{\ B'\mu'\nu'} \sigma^{\nu'} 
+ O(\epsilon^2). 
\label{gradUAB2} 
\end{equation} 
Additional differentiations produce
\begin{subequations} 
\label{gradUAB3} 
\begin{align} 
\nabla_{\mu'} \nabla_{\nu'} U^A_{\ B'} &= g^A_{\ A'} \Bigl( 
\nabla_{\mu'} U^{A'}_{\ B'\nu'} + \nabla_{\nu'} U^{A'}_{\ B'\mu'}  
+ U^{A'}_{\ B'\mu'\nu'} \Bigr) 
- \frac{1}{2} R^A_{\ B'\mu'\nu'} +O(\epsilon), \\ 
\nabla_{\mu} \nabla_{\nu'} U^A_{\ B'} &= 
-g^{\mu'}_{\ \mu} g^A_{\ A'} \Bigl( \nabla_{\nu'} U^{A'}_{\ B'\mu'}   
+ U^{A'}_{\ B'\mu'\nu'} \Bigr) 
+ \frac{1}{2} g^{\mu'}_{\ \mu} R^A_{\ B'\mu'\nu'} + O(\epsilon), \\ 
\nabla_{\mu} \nabla_{\nu} U^A_{\ B'} &= 
g^{\mu'}_{\ \mu} g^{\nu'}_{\ \nu} g^A_{\ A'} U^{A'}_{\ B'\mu'\nu'}
+ \frac{1}{2} g^{\mu'}_{\ \mu} g^{\nu'}_{\ \nu} R^A_{\ B'\mu'\nu'} 
+ O(\epsilon). 
\end{align} 
\end{subequations} 

The leading term of an expansion of $V^A_{\ B'}$ in powers of
$\sigma^{\mu'}$ can readily be obtained by inserting 
\begin{equation} 
V^A_{\ B'}(x,x') = g^A_{\ A'} \Bigl[ V^{A'}_{\ B'} 
+ O(\epsilon) \Bigr] 
\label{V_taylor1}
\end{equation}  
on the left-hand side of Eq.~(\ref{V_deq1}) and substituting
Eqs.~(\ref{U_taylor1}), (\ref{gradUAB2}), and (\ref{gradUAB3}) on the
right-hand side. After involvement of Eq.~(\ref{U_taylor2}) and some
simplification, we arrive at 
\begin{equation} 
V^{A'}_{\ B'} = -\frac{1}{4} \nabla^{\mu'} M^{A'}_{\ B'\mu'} 
- \frac{1}{8} M^{A'}_{\ C'\mu'} M^{C'\ \mu'}_{\ B'} 
+ \frac{1}{2} N^{A'}_{\ B'} 
+ \frac{1}{12} \delta^{A'}_{\ B'} R', 
\label{V_taylor2} 
\end{equation} 
in which $N^{A'}_{\ B'}$ is the tensor field $N^A_{\ B}$ evaluated at
$x'$, and $R'$ is the Ricci scalar at $x'$.   

\section{Potentials near the world line} 
\label{sec:potentials} 

\subsection{Retarded, singular, and regular potentials} 

With $\psi^A = \{ \gamma^{\alpha\beta}, b^\alpha, \phi \}$, the
solutions to Eqs.~(\ref{wave_grav_scalar}), (\ref{wave_scalar}), and
(\ref{wave_em}) are all given by the expression of Eq.~(\ref{psi_sol2}), 
\begin{equation} 
\psi^A(x) = \int_\gamma G^A_{\ M}(x,z) q^M(\tau)\, d\tau,  
\label{psi_sol} 
\end{equation} 
in which $q^M = \{ 4(m-q\Phi)u^\mu u^\nu, e u^\mu, q \}$; summation
over the repeated meta-index includes all individual indices (or
absence of index) contained in $M$.    

Following techniques detailed in Sec.~17.2 of PPV, we find that the
retarded solution to the perturbation equation can be expressed as 
\begin{equation} 
\psi^A(x) = \frac{1}{r} U^A_{\ B'}(x,x') q^{B'}(u) 
+ \int_{\tau_<}^u V^A_{\ M}(x,z) q^M(\tau)\, d\tau
+ \int_{-\infty}^{\tau_<} G^A_{\ M}(x,z) q^M(\tau)\, d\tau 
\label{psi_ret} 
\end{equation} 
when $x$ is close to the world line $z(\tau)$. Here $x' = z(u)$
denotes the retarded point associated with $x$, defined by the
null condition $\sigma(x,x') = 0$, 
$r := \sigma_{\alpha'}(x,x') u^{\alpha'}$ is the retarded distance to
the world line, and $\tau_<$ is the proper time at which the world
line enters the normal convex neighborhood of the field point $x$. 

Following techniques detailed in Sec.~17.5 of PPV, we identify the
Detweiler-Whiting singular potential as 
\begin{equation} 
\psi^A_{\sf S}(x) = \frac{1}{2r} U^A_{\ B'}(x,x') q^{B'}(u) 
+ \frac{1}{2r_{\rm adv}} U^A_{\ B''}(x,x'') q^{B''}(v) 
- \frac{1}{2} \int_u^v V^A_{\ M}(x,z) q^M(\tau)\, d\tau, 
\label{psi_sing} 
\end{equation} 
where $x'' = z(v)$ is the advanced point associated with $x$, and
$r_{\rm adv} := -\sigma_{\alpha''} u^{\alpha''}$ is the advanced
distance to the world line. This also is a solution to the
inhomogeneous wave equation, as written in Eq.~(\ref{metaFE}).  
The Detweiler-Whiting regular potential is then $\psi^A_{\sf R} :=  
\psi^A - \psi^A_{\sf S}$, or
\begin{align} 
\psi^A_{\sf R}(x) &= \frac{1}{2r} U^A_{\ B'}(x,x') q^{B'}(u) 
- \frac{1}{2r_{\rm adv}} U^A_{\ B''}(x,x'') q^{B''}(v) 
+ \int_{\tau_<}^u V^A_{\ M}(x,z) q^M(\tau)\, d\tau
\nonumber \\ & \quad \mbox{} 
+ \frac{1}{2} \int_u^v V^A_{\ M}(x,z) q^M(\tau)\, d\tau 
+ \int_{-\infty}^{\tau_<} G^A_{\ M}(x,z) q^M(\tau)\, d\tau. 
\label{psi_reg} 
\end{align} 
 
\subsection{Regular potential in Fermi coordinates} 

The regular potential $\psi^A_{\sf R}$ will be implicated in the 
equations of motion satisfied by the particle, and for the purpose of
deriving these equations it is convenient to obtain explicit
expressions in terms of Fermi coordinates. The construction of these  
coordinates is detailed in Sec.~9 of PPV. 

The Fermi coordinates $(t,x^a)$ refer to a point $\bar{x} = z(t)$ on
the world line which is simultaneous to $x$, in the sense that $x$
and $\bar{x}$ are linked by a spacelike geodesic that is orthogonal to
the world line. The precise condition is $\sigma_{\bar{\alpha}}
u^{\bar{\alpha}} = 0$, and the geodesic distance $s$ from $\bar{x}$ to
$x$ is given by $s^2 = 2\sigma(\bar{x},x)$. The spatial Fermi
coordinates are defined by $x^a := -\sigma^{\bar{\alpha}}
e^a_{\bar{\alpha}}$, in which $e^{\bar{\alpha}}_a$ is a triad of
Fermi-Walker transported spatial vectors on the world line (triad
indices are raised with $\delta^{ab}$). From this it follows that 
$\sigma^{\bar{\alpha}} = -x^a e^{\bar{\alpha}}_a$.  

We must relate the retarded point $x' = z(u)$ and the advanced point
$x'' = z(v)$ to the simultaneous point $\bar{x} = z(t)$. To achieve
this we rely on expansion techniques developed in Sec.~11 of PPV. We
introduce $\Delta := t-u$, $\Delta' := v-t$, as well as the scalar
function  
\begin{equation} 
\sigma(\tau) := \sigma\bigl( x, z(\tau) \bigr) 
\end{equation} 
on the world line; $x$ is taken to be fixed on the right-hand side of
this relation. In this notation we have that $\sigma(u) = 0$,
$\sigma(t) = \frac{1}{2} s^2$, and $\sigma(v) = 0$. To obtain $\Delta$
we write $0 = \sigma(u) = \sigma(t - \Delta)$, expand the
right-hand side in powers of $\Delta$, and solve for $\Delta$
expressed as an expansion in powers of $s$. To obtain $\Delta'$ we
start instead with $0 = \sigma(v) = \sigma(t + \Delta')$.  In this way
we obtain 
\begin{subequations} 
\label{Delta2} 
\begin{align} 
\Delta &= s \biggl\{ 1 - \frac{1}{2} a_a x^a 
+ \frac{3}{8} (a_a x^a)^2 
+ \frac{1}{24} \dot{a}_t s^2  
+ \frac{1}{6} s \dot{a}_a x^a 
- \frac{1}{6} R_{tatb} x^a x^b + O(s^3)  \biggr\}, \\
\Delta' &= s \biggl\{ 1 - \frac{1}{2} a_a x^a 
+ \frac{3}{8} (a_a x^a)^2 
+ \frac{1}{24} \dot{a}_t s^2  
- \frac{1}{6} s \dot{a}_a x^a 
- \frac{1}{6} R_{tatb} x^a x^b + O(s^3) \biggr\},   
\end{align} 
\end{subequations} 
where $a_a := a_{\bar{\alpha}} e^{\bar{\alpha}}_a$, $R_{tatb} :=
R_{\bar{\gamma}\bar{\alpha}\bar{\delta}\bar{\beta}} u^{\bar{\gamma}} 
e^{\bar{\alpha}}_a u^{\bar{\delta}} e^{\bar{\beta}}_b$ and so on are
components of tensors in Fermi coordinates, evaluated at $\bar{x}$. 

To relate $r$ to $s$ we notice that 
$r = \dot{\sigma}(u) = \dot{\sigma}(t-\Delta)$, which can be
expanded in powers of $\Delta$. Similarly we have that
$r_{\rm adv} = -\dot{\sigma}(v) = -\dot{\sigma}(t+\Delta')$, which can
be expanded in powers of $\Delta'$. With the expressions provided in 
Eq.~\ref{Delta2}, we eventually obtain 
\begin{subequations} 
\label{Delta3} 
\begin{align} 
r &= s \biggl\{ 1 + \frac{1}{2} a_a x^a 
- \frac{1}{8} (a_a x^a)^2 
- \frac{1}{8} \dot{a}_t s^2  
- \frac{1}{3} s \dot{a}_a x^a 
+ \frac{1}{6} R_{tatb} x^a x^b + O(s^3) \biggr\}, \\
r_{\rm adv} &= s \biggl\{ 1 + \frac{1}{2} a_a x^a 
- \frac{1}{8} (a_a x^a)^2 
- \frac{1}{8} \dot{a}_t s^2  
+ \frac{1}{3} s \dot{a}_a x^a 
+ \frac{1}{6} R_{tatb} x^a x^b + O(s^3) \biggr\}. 
\end{align}
\end{subequations} 
From all this we can form the combinations
\begin{subequations} 
\label{Delta4} 
\begin{align} 
\frac{1}{2} \biggl( \frac{1}{r} - \frac{1}{r_{\rm adv}} \biggr) 
&= \frac{1}{3} \dot{a}_a x^a + O(s^2), \\ 
\frac{1}{2} \biggl( \frac{\Delta}{r} +\frac{\Delta'}{r_{\rm adv}} \biggr) 
&= 1 - a_a x^a + O(s^2), \\
\frac{1}{2} \biggl( \frac{\Delta^2}{r} 
- \frac{\Delta^{\prime 2}}{r_{\rm adv}} \biggr)  
&= O(s^3), 
\end{align} 
\end{subequations} 
which will be required in a moment. 

To express the regular potential in Fermi coordinates we must relate
$U^A_{\ B'} q^{B'}$ and $U^A_{\ B''} q^{B''}$ to quantities defined at
the simultaneous point $\bar{x}$. To achieve this we define   
\begin{equation} 
U^A(\tau) := U^A_{\ M}\bigl( x, z(\tau) \bigr) q^M(\tau),  
\label{UA_def} 
\end{equation}  
write $U^A_{\ B'} q^{B'} = U^A(u) = U^A(t-\Delta)$ and  
$U^A_{\ B''} q^{B''} = U^A(v) = U^A(t+\Delta')$, and expand in powers
of $\Delta$ or $\Delta'$. The derivatives of $U^A(\tau)$ can be
calculated with the help of Eqs.~(\ref{gradUAB1}) and
(\ref{gradUAB3}). We have  
\begin{subequations} 
\label{UA_derivs} 
\begin{align} 
U^A &= g^A_{\ \bar{A}} \Bigl[ q^{\bar{A}} 
+ q^{\bar{B}} U^{\bar{A}}_{\ \bar{B} \bar{\mu}} \sigma^{\bar{\mu}} 
+ \frac{1}{2} q^{\bar{B}} U^{\bar{A}}_{\ \bar{B} \bar{\mu} \bar{\nu}} 
   \sigma^{\bar{\mu}} \sigma^{\bar{\nu}} \Bigl] + O(s^3), \\ 
\dot{U}^A &= g^A_{\ \bar{A}} \biggl[ \dot{q}^{\bar{A}} 
+ q^{\bar{B}} U^{\bar{A}}_{\ \bar{B} \bar{\mu}} u^{\bar{\mu}} 
+ \Bigl( q^{\bar{B}} \dot{U}^{\bar{A}}_{\ \bar{B} \bar{\nu}} 
+ \dot{q}^{\bar{B}} U^{\bar{A}}_{\ \bar{B} \bar{\nu}}
+ q^{\bar{B}} U^{\bar{A}}_{\ \bar{B} \bar{\mu} \bar{\nu}}
   u^{\bar{\mu}} \Bigr) \sigma^{\bar{\nu}} \biggr] 
+ \frac{1}{2} q^{\bar{B}} R^A_{\ \bar{B}\bar{\mu}\bar{\nu}}
   u^{\bar{\mu}} \sigma^{\bar{\nu}} + O(s^2), \\ 
\ddot{U}^A &= g^A_{\ \bar{A}} \Bigl[  \ddot{q}^{\bar{A}} 
+ 2 \dot{q}^{\bar{B}} U^{\bar{A}}_{\ \bar{B}\bar{\mu}} u^{\bar{\mu}} 
+ 2 q^{\bar{B}} \dot{U}^{\bar{A}}_{\ \bar{B}\bar{\mu}} u^{\bar{\mu}} 
+ q^{\bar{B}} U^{\bar{A}}_{\ \bar{B}\bar{\mu}} a^{\bar{\mu}} 
+ q^{\bar{B}} U^{\bar{A}}_{\ \bar{B}\bar{\mu}\bar{\nu}} 
   u^{\bar{\mu}} u^{\bar{\nu}} \Bigr] + O(s), 
\end{align} 
\end{subequations}
where $\dot{q}^A$ and $\ddot{q}^A$ were introduced in
Eqs.~(\ref{qdot_def}) and (\ref{qddot_def}), respectively, 
and $\dot{U}^M_{\ N\mu} := u^\nu \nabla_\nu  
U^M_{\ N\mu}$; the barred tensorial indices indicate that the
expressions are evaluated at $\tau = t$.    

Defining 
\begin{equation} 
V^A(\tau) := V^A_{\ M}\bigl( x, z(\tau) \bigr) q^M(\tau) 
\label{VA_def} 
\end{equation}  
the integrals involving $V^A_{\ M} q^M$ in Eq.~(\ref{psi_reg}) can be  
written as 
\begin{equation} 
\int_{\tau_<}^u V^A_{\ M} q^M\, d\tau
+ \frac{1}{2} \int_u^v V^A_{\ M} q^M\, d\tau 
= \int_{\tau_<}^t V^A\, d\tau
- \int_u^t V^A\, d\tau
+ \frac{1}{2} \int_u^v V^A\, d\tau,  
\end{equation} 
and close to the world line the last two terms evaluate to
$-\frac{1}{2} (\Delta-\Delta') V^A(t) = O(s^2)$. For future reference
we note that 
\begin{equation} 
V^A(t) = g^A_{\ \bar{A}} V^{\bar{A}}_{\ \bar{B}} q^{\bar{B}} 
+ O(s). 
\label{VA_fermi} 
\end{equation} 

Making the substitutions in Eq.~(\ref{psi_reg}) produces  
\begin{subequations}
\label{psi_fermi} 
\begin{align} 
 \psi^A_{\sf R}(t,x^a) &= -(1-a_c x^c) \dot{U}^A(t) 
+ \frac{1}{3} U^A(t)\, \dot{a}_c x^c 
+ \psi^A[\text{tail}] + O(s^2) 
\label{psi_fermi_a} \\ 
&= -g^A_{\ \bar{A}} \Bigl( \dot{q}^{\bar{A}} 
+ q^{\bar{B}} U^{\bar{A}}_{\ \bar{B} t} \Bigr) (1-a_c x^c) 
+ g^A_{\ \bar{A}} \biggl( \frac{1}{3} q^{\bar{A}} \dot{a}_c 
+ q^{\bar{B}} \dot{U}^{\bar{A}}_{\ \bar{B} c} 
+ \dot{q}^{\bar{B}} U^{\bar{A}}_{\ \bar{B} c}  
+ q^{\bar{B}} U^{\bar{A}}_{\ \bar{B} tc} \biggr) x^c 
\nonumber \\ & \quad \mbox{} 
+ \frac{1}{2} q^{\bar{B}} R^A_{\ \bar{B} tc}\, x^c 
+ \psi^A[\text{tail}] + O(s^2), 
\label{psi_fermi_b} 
\end{align} 
\end{subequations}  
where  
\begin{subequations}
\label{psi_tail} 
\begin{align} 
 \psi^A[\text{tail}](x) &:= 
\int_{\tau_<}^t V^A_{\ M}(x,z) q^M(\tau)\,  d\tau 
+ \int_{-\infty}^{\tau_<} G^A_{\ M}(x,z) q^M(\tau)\,  d\tau \\
&= \int_{-\infty}^{t^-} G^A_{\ M}(x,z) q^M(\tau)\,  d\tau 
\end{align} 
\end{subequations} 
is the ``tail part'' of the potential; in the second form the
integration is cut short at $\tau = t^- := t - 0^+$ to avoid the 
singular behavior of the Green's function when the limit 
$x \to \bar{x}$ is eventually taken.

A complete listing of the potentials $\psi^A_{\sf R} 
= \{\gamma^{\alpha\beta}_{\sf R}, b^\alpha_{\sf R}, \phi_{\sf R}\}$
can now be obtained from Eq.~(\ref{psi_fermi_b}). The calculation
requires explicit expressions for the tensors 
$U^{\bar{A}}_{\ \bar{B}\bar{\mu}}$, 
$U^{\bar{A}}_{\ \bar{B}\bar{\mu}\bar{\nu}}$, and
$V^{\bar{A}}_{\ \bar{B}}$, and these will be computed
at a later stage. It also requires the components of the parallel 
propagator $g^{\alpha}_{\ \bar{\beta}}$ in Fermi coordinates. These
are (PPV Sec.~9.4)  
\begin{equation} 
g^t_{\ \bar{t}} = 1 - a_a x^a + O(s^2), \qquad  
g^t_{\ \bar{a}} =  O(s^2), \qquad  
g^a_{\ \bar{t}} = O(s^2), \qquad  
g^a_{\ \bar{b}} = \delta^a_{\ b} + O(s^2). 
\label{parallel-propagator}   
\end{equation} 

\subsection{Singular potential in covariant form} 
\label{subsec:sing_general} 

The singular potentials $\psi^A_{\sf S}$ are often implicated in the
calculation of regularization parameters or regularized sources for
self-force computations. The starting point of such calculations 
is a covariant expression for the singular potentials, which can then
be evaluated in any convenient coordinate system and involved in
computations of effective sources or regularization parameters. We
provide such a covariant expression here, using methods first devised
by Haas and Poisson (2006).   

The singular potential is defined by Eq.~(\ref{psi_sing}), and we wish
to consolidate its dependence on the world line from the two points
$x'$ and $x''$ related to $x$ by a null condition, to a single point
$\hat{x}$ that has no particular relationship to $x$; the points $x$
and $\hat{x}$ are assumed to be in a spacelike relation, but otherwise
$\hat{x}$ is a fixed arbitrary point on the world line. We let 
$\hat{x} = z(\hat{\tau})$, $\Delta_- := u - \hat{\tau}$, 
$\Delta_+ := v - \hat{\tau}$, and 
\begin{equation} 
\hat{r} := \sigma_{\hat{\alpha}}(x,\hat{x})
u^{\hat{\alpha}}(\hat{\tau}), \qquad 
\rho^2 := \bigl( g^{\hat{\alpha}\hat{\beta}} + u^{\hat{\alpha}}
u^{\hat{\beta}} \bigr) \sigma_{\hat{\alpha}}(x,\hat{x})
\sigma_{\hat{\beta}}(x,\hat{x}). 
\label{rho_def} 
\end{equation}  
To express $\psi^A_{\sf S}$ in terms of $\hat{x}$ we apply the same
methods as in the preceding subsection. First, $\Delta_\pm$ is
determined by writing $0 = \sigma(u) = \sigma(\hat{\tau} + \Delta_-)$
or $0 = \sigma(v) = \sigma(\hat{\tau} + \Delta_+)$ and expanding in 
powers of $\Delta_\pm$; here $\sigma(\tau) := \sigma(x,z(\tau))$. With 
$\sigma(\hat{\tau}) = \frac{1}{2}(\rho^2 - \hat{r}^2)$,
$\dot{\sigma}(\hat{\tau}) = \hat{r}$, and so on, we obtain an
expansion of the form 
\begin{equation} 
\Delta_\pm = (\hat{r} \pm \rho) 
\pm \frac{(\hat{r} \pm \rho)^2}{2\rho} a_{\hat{\alpha}}
\sigma^{\hat{\alpha}} + \cdots, 
\end{equation} 
in which the first group of terms is of order $\rho$, the second group
is of order $\rho^2$, and a third group of terms at order $\rho^3$
was calculated but is too large to be displayed here. Second, $r$ and
$r_{\rm adv}$ are related to $\rho$ through the relations 
$r = \dot{\sigma}(u) = \dot{\sigma}(\hat{\tau} + \Delta_-)$
and $r_{\rm adv} = -\dot{\sigma}(v) = -\dot{\sigma}(\hat{\tau} 
+ \Delta_+)$, which are also expanded in powers of
$\Delta_\pm$. Third, $U^A_{\ B'} q^{B'} = U^A(u)$ and 
$U^A_{\ B''} q^{B''} = U^A(v)$ are expressed in a similar way, 
and fourth, the integral of $V^A := V^A_{\ M} q^M$ is evaluated as  
$-\frac{1}{2} (\Delta_+-\Delta_-) V^A(\hat{\tau})$. 

Collecting results, we obtain the covariant expression 
\begin{equation} 
\psi^A_{\sf S} = \frac{1}{\rho} \biggl[ \gamma_1 U^A(\hat{\tau}) 
+ \hat{r} \gamma_2 \dot{U}^A(\hat{\tau}) 
+ \frac{1}{2} (\rho^2 + \hat{r}^2) \ddot{U}^A(\hat{\tau}) 
- \rho^2 V^A(\hat{\tau}) + O(\rho^3) \biggr] 
\label{psi_sing_cov} 
\end{equation} 
for the singular potential. We have introduced
\begin{subequations} 
\begin{align} 
\gamma_1 &:= \frac{\rho}{2} \biggl(\frac{1}{r} 
+ \frac{1}{r_{\rm adv}} \biggr)\\ 
&= 1 + \frac{\rho^2 - \hat{r}^2}{2\rho^2} 
a_{\hat{\mu}} \sigma^{\hat{\mu}} 
- \frac{3\rho^4 + 6\rho^2\hat{r}^2 - \hat{r}^4}{24\rho^2} a^2 
+ \frac{3(\rho^2-\hat{r}^2)^2}{8\rho^4} 
\bigl( a_{\hat{\mu}} \sigma^{\hat{\mu}} \bigr)^2 
+ \frac{\hat{r}(3\rho^2-\hat{r}^2)}{6\rho^2}  
\dot{a}_{\hat{\mu}} \sigma^{\hat{\mu}}  
\nonumber \\ & \quad \mbox{}
- \frac{\rho^2-\hat{r}^2}{6\rho^2}
R_{\hat{\mu}\hat{\lambda}\hat{\nu}\hat{\rho}} 
u^{\hat{\mu}} \sigma^{\hat{\lambda}}  
u^{\hat{\nu}} \sigma^{\hat{\rho}}
+ O(\rho^3),  
\end{align} 
\end{subequations} 
where $a^2 := a_{\hat{\mu}} a^{\hat{\mu}}$, and 
\begin{equation} 
\gamma_2 := \frac{\rho}{2 \hat{r}} \biggl(\frac{\Delta_-}{r}  
+ \frac{\Delta_+}{r_{\rm adv}} \biggr) 
= 1 + \frac{3\rho^2-\hat{r}^2}{2\rho^2} 
a_{\hat{\mu}} \sigma^{\hat{\mu}} + O(\rho^2).   
\end{equation}  
The tensors $U^A$, $\dot{U}^A$, $\ddot{U}^A$ can be imported from 
Eq.~(\ref{UA_derivs}), and $V^A$ can be obtained from
Eq.~(\ref{VA_fermi}), with the understanding that these objects are
now evaluated at $\tau = \hat{\tau}$ instead of $\tau = t$, so that
the expressions involve tensors with hatted indices instead of barred 
indices. Explicit expressions will be given below. 

\section{Self-force in scalarvac spacetimes} 
\label{sec:scalar_sf} 

In this section we compute the self-force acting on a particle of mass
$m$ and scalar charge $q$ in a scalarvac spacetime; the metric of the
background spacetime is $g_{\alpha\beta}$, and the background scalar
field is $\Phi$. To simplify the calculations we adopt the gauge of
Eq.~(\ref{lambda-gauge}) with $\lambda = 1$. A glance at
Eq.~(\ref{MN-scalar1}) reveals that this choice of gauge eliminates
the term $M^A_{\ B\mu} \nabla^\mu \psi^B$ from the perturbation
equation, and this considerably simplifies the structure of the
Hadamard Green's function. 

\subsection{Equations of motion} 

On a formal level, the particle's equations of motion are formulated
in the perturbed spacetime. We have 
\begin{equation} 
{\sf m} \frac{ {\sf D} {\sf u}^\mu }{d {\sf t}} 
= q \bigl( {\sf g}^{\mu\nu} + {\sf u}^\mu {\sf u}^{\nu} \bigr) 
{\sf D}_\mu {\sf \Phi},  
\end{equation} 
where ${\sf m} := m - q {\sf \Phi}$ is an inertial mass parameter that
satisfies 
\begin{equation} 
\frac{d {\sf m}}{d {\sf t}} = -q {\sf u}^\mu {\sf D}_{\mu} {\sf \Phi}. 
\end{equation} 
We have introduced $\sf t$ as proper time in the perturbed spacetime, 
${\sf u}^\mu = dz^\mu/d{\sf t}$, and ${\sf D}/d{\sf t}$ indicates
covariant differentiation (in the perturbed spacetime) along the world
line.   

Substitution of ${\sf g}_{\alpha\beta} = g_{\alpha\beta} +
h_{\alpha\beta}$, ${\sf \Phi} = \Phi + \phi$, $d\tau/d{\sf t} = 1 +
\frac{1}{2} h_{\mu\nu} u^\mu u^\nu$, and linearization with respect to
all perturbations produce  
\begin{equation} 
{\sf m} a^\mu = \bigl( g^{\mu\nu} + u^\mu u^\nu \bigr) \Bigl[ 
q \nabla_\nu \Phi - q u^\lambda u^\rho h_{\lambda\rho} \nabla_\nu \Phi  
- q h^\lambda_{\ \nu} \nabla_\lambda \Phi 
+ q \nabla_\nu \phi 
- \frac{1}{2} {\sf m} \bigl( 2 \nabla_\rho h_{\nu\lambda} 
- \nabla_\nu h_{\lambda\rho} \bigr) u^\lambda u^\rho \Bigr]
\label{sf_grav_sca1} 
\end{equation} 
and 
\begin{equation} 
\frac{d {\sf m}}{d\tau} = -q u^\mu \nabla_\mu \Phi 
- q u^\mu \nabla_\mu \phi, 
\label{sf_grav_sca2} 
\end{equation} 
in which all quantities refer to the background spacetime. 

As was mentioned, these equations have formal validity only, because
the retarded potentials $h_{\alpha\beta}$ and $\phi$ are singular on
the world line. To make sense of these equations we follow the
Detweiler-Whiting prescription \cite{detweiler-whiting:03}, which
asserts that the motion of the particle is driven entirely by the
regular piece of the potentials, obtained after removal of the
singular piece:  
$h^{\alpha\beta}_{\sf R} = h^{\alpha\beta} - h^{\alpha\beta}_{\sf S}$
and $\phi_{\sf R} = \phi - \phi_{\sf S}$. The equations of motion,
therefore, are written as in Eqs.~(\ref{sf_grav_sca1}) and
(\ref{sf_grav_sca2}), but with the regular potentials standing in for
the retarded potentials. 

In Fermi coordinates we have 
\begin{equation} 
{\sf m} a_a = q \nabla_a \Phi - q h_{tt} \nabla_a \Phi 
- q h^t_{\ a} \nabla_t \Phi - q h^b_{\ a} \nabla_b \Phi 
+ q \nabla_a \phi - \frac{1}{2} {\sf m} \bigl( 2 \nabla_t h_{ta} 
- \nabla_a h_{tt} \bigr)  
\label{sf_grav_sca3} 
\end{equation} 
and 
\begin{equation} 
\frac{d {\sf m}}{dt} = -q \nabla_t \Phi - q \nabla_t \phi,  
\label{sf_grav_sca4} 
\end{equation} 
where we suppress the label ``${\sf R}$'' on the potentials
$h_{\alpha\beta}$ and $\phi$.

\subsection{Regular potentials} 

Importing $M^A_{\ B\mu}$ and $N^A_{\ B}$ from
Sec.~\ref{sec:scalarvac} and inserting them within
Eqs.~(\ref{U_taylor2}) and (\ref{V_taylor2}) reveals that the tensors
that appear in the Hadamard Green's function are given explicitly by  
\begin{subequations} 
\label{grav_scalar_tensorlist} 
\begin{align} 
U^A_{\ B\mu} &= 0, \\ 
U^{\alpha\beta}_{\ \ \ |\gamma\delta\mu\nu} &= 
\frac{1}{6} \delta^{(\alpha}_{\ \gamma} \delta^{\beta)}_{\ \delta} 
R_{\mu\nu}, \\ 
U^{\cdot}_{\ |\cdot\mu\nu} &= \frac{1}{6} R_{\mu\nu}, \\ 
V^{\alpha\beta}_{\ \ \ |\gamma\delta} &= 
R^{\alpha\ \, \beta}_{\ (\gamma\ \delta)} 
- \delta^{\alpha}_{\ (\gamma} \nabla^\beta \Phi \nabla_{\delta)} \Phi 
+ \frac{1}{12} \delta^{\alpha}_{\ (\gamma} \delta^{\beta}_{\ \delta)}
    R, \\
V^{\alpha\beta}_{\ \ \ |\cdot} &= -2 \nabla^\alpha \nabla^\beta \Phi, \\ 
V^\cdot_{\ |\alpha\beta} &= -\frac{1}{2} \biggl( 
  \nabla_\alpha \nabla_\beta \Phi - \frac{1}{2} F' g_{\alpha\beta}
  \biggr), \\ 
V^\cdot_{\ |\cdot} &= -\frac{1}{2} \Bigl( 
  2 \nabla_\mu \Phi \nabla^\mu \Phi + F'' \Bigr) 
+ \frac{1}{12} R, 
\end{align} 
\end{subequations} 
where we omit the primes on indices to keep the notation uncluttered.  

Making the substitutions in Eq.~(\ref{psi_fermi_b}), we obtain 
\begin{subequations} 
\begin{align} 
\gamma^{tt}_{\sf R} &= 4 q \dot{\Phi} (1 - 3 a_c x^c) 
+ \frac{2}{3} (m-q\Phi) \bigl( 2 \dot{a}_c + R_{tc}
\bigr) x^c + \gamma^{tt}[\text{tail}] + O(s^2), \\ 
\gamma^{ta}_{\sf R} &= -4(m-q\Phi) a^a 
+ 2(m-q\Phi) \bigl( 4 a^a a_c - R^a_{\ tct} \bigr) x^c 
+ \gamma^{ta}[\text{tail}] + O(s^2), \\ 
\gamma^{ab}_{\sf R} &= \gamma^{ab}[\text{tail}] + O(s^2) 
\end{align} 
\end{subequations} 
for the gravitational potentials, and 
\begin{equation} 
\phi_{\sf R} = \frac{1}{6} q (2 \dot{a}_a + R_{ta}) x^c
+ \phi[\text{tail}] + O(s^2) 
\label{phi_reg} 
\end{equation} 
for the scalar perturbation. The tail terms are 
\begin{equation} 
\gamma^{\alpha\beta}[\text{tail}] := 
4\int_{-\infty}^{t^-} (m-q\Phi) G^{\alpha\beta}_{\ \ \ |\mu\nu}(x,z) 
u^\mu u^\nu\, d\tau 
+ q \int_{-\infty}^{t^-} G^{\alpha\beta}_{\ \ \ |\cdot}(x,z)\, d\tau 
\end{equation} 
and 
\begin{equation} 
\phi[\text{tail}] := 
4\int_{-\infty}^{t^-} (m-q\Phi) G^{\cdot}_{\ |\mu\nu}(x,z) 
u^\mu u^\nu\, d\tau 
+ q \int_{-\infty}^{t^-} G^{\cdot}_{\ |\cdot}(x,z)\, d\tau. 
\label{scalar_phi_tail} 
\end{equation}
After trace reversal and lowering the indices using the metric 
$g_{tt} = -1 - 2a_c x^c + O(s^2)$, $g_{ta} = O(s^2)$, 
$g_{ab} = \delta_{ab} + O(s^2)$, the gravitational potentials become 
\begin{subequations} 
\begin{align} 
h_{tt}^{\sf R} &= 2 q \dot{\Phi}( 1 + a_c x^c ) 
+ \frac{1}{3} (m-q\Phi) \bigl( 2 \dot{a}_c + R_{tc}
\bigr) x^c+ h_{tt}[\text{tail}] + O(s^2), \\ 
h_{ta}^{\sf R} &= 4(m-q\Phi) a_a 
+ 2(m-q\Phi) R_{atct} x^c 
+ h_{ta}[\text{tail}] + O(s^2), \\ 
h_{ab}^{\sf R} &= 2 q \dot{\Phi} \delta_{ab} (1-a_c x^c) 
+ \frac{1}{3} (m-q\Phi) \delta_{ab} 
\bigl( 2 \dot{a}_c + R_{tc} \bigr) x^c 
+ h_{ab}[\text{tail}] + O(s^2),  
\end{align} 
\end{subequations} 
with 
\begin{equation} 
h_{\alpha\beta}[\text{tail}] := 
4 \int_{-\infty}^{t^-} (m-q\Phi)\bar{G}_{\alpha\beta|\mu\nu}(x,z) 
u^\mu u^\nu\, d\tau 
+ q \int_{-\infty}^{t^-} \bar{G}_{\alpha\beta|\cdot}(x,z)\, d\tau,
\label{scalar_h_tail} 
\end{equation} 
where the overbar indicates the operation of trace reversal. 

The relevant covariant derivatives of the potentials are 
\begin{subequations} 
\begin{align} 
\nabla_t h_{ta} &= 4(m-q\Phi) \bigl( \dot{a}_a 
+ \bar{V}_{ta|tt} \bigr) - 8q \dot{\Phi} a_a  
+ q \bar{V}_{ta|\cdot} + h_{tat}[\text{tail}] + O(s), \\ 
\nabla_a h_{tt} &= \frac{1}{3} (m-q\Phi) \bigl( 2\dot{a}_a 
+ R_{ta}\bigr)  + h_{tta}[\text{tail}] + O(s), \\ 
\nabla_t \phi &= q V^\cdot_{\ |\cdot} 
+ 4 (m-q\Phi) V^\cdot_{\ |tt} + \phi_t[\text{tail}] + O(s), \\
\nabla_a \phi &= \frac{1}{6} q \bigl( 2\dot{a}_a + R_{ta} \bigr) 
+ \phi_a[\text{tail}] + O(s), 
\end{align} 
\end{subequations}  
where 
\begin{subequations}
\label{scalar_grad_tail} 
\begin{align} 
h_{\alpha\beta\gamma}[\text{tail}] &:= 
4\int_{-\infty}^{t^-} (m-q\Phi) \nabla_\gamma 
\bar{G}_{\alpha\beta|\mu\nu}(x,z) u^\mu u^\nu\, d\tau 
+ q \int_{-\infty}^{t^-} \nabla_\gamma 
\bar{G}_{\alpha\beta|\cdot}(x,z)\, d\tau, \\
\phi_\alpha[\text{tail}] &:= 
4\int_{-\infty}^{t^-} (m-q\Phi) \nabla_\alpha 
G^{\cdot}_{\ |\mu\nu}(x,z) u^\mu(\tau) u^\nu(\tau)\, d\tau 
+ q \int_{-\infty}^{t^-} \nabla_\alpha 
G^{\cdot}_{\ |\cdot}(x,z)\, d\tau. 
\end{align} 
\end{subequations} 
To arrive at these results we relied on the fact that
in Fermi coordinates, the relevant Christoffel symbols are 
$\Gamma^t_{t a} = a_a + O(s)$ and $\Gamma^a_{tt} = a^a + O(s)$; all
other symbols are $O(s)$ and not required in this computation. 

\subsection{Explicit form of the equations of motion} 

Making the substitutions in Eqs.~(\ref{sf_grav_sca3}) and 
(\ref{sf_grav_sca4}), we obtain 
\begin{align} 
{\sf m} a_a &= q \nabla_a \Phi 
+ {\sf m}^2 \biggl( -\frac{11}{3} \dot{a}_a 
+ \frac{1}{6} R_{ta} - 2 \nabla_t \Phi \nabla_a \Phi \biggr) 
+ q^2 \biggl( \frac{1}{3} \dot{a}_a + \frac{1}{6} R_{ta} 
- 4 \nabla_t \Phi \nabla_a \Phi \biggr) 
\nonumber \\ & \quad \mbox{}
+ {\sf m} q \Bigl( 12 a_a \nabla_t \Phi  
+ 2 \nabla_t \nabla_a \Phi \Bigr)  
- q h_{tt}[\text{tail}] \nabla_a \Phi 
- q h^t_{\ a}[\text{tail}] \nabla_t \Phi -
 q h^b_{\ a}[\text{tail}] \nabla_b \Phi 
\nonumber \\ & \quad \mbox{}
+ q \phi_a [\text{tail}] 
- \frac{1}{2} {\sf m} \bigl( 2 h_{tat}[\text{tail}] 
- h_{tta}[\text{tail}] \bigr)
\label{eom_scalar1} 
\end{align} 
and 
\begin{equation} 
\frac{d {\sf m}}{dt} = -q \nabla_t \Phi 
+ q^2 \biggl( \nabla_\mu \Phi \nabla^\mu \Phi 
+ \frac{1}{2} F'' - \frac{1}{12} R \biggr) 
+ {\sf m} q \Bigl( 2 \nabla_t \nabla_t \Phi + F' \Bigr) 
- q \phi_t[\text{tail}].  
\label{eom_scalar2} 
\end{equation} 
In the perturbation terms we no longer distinguish between $m-q\Phi$
and ${\sf m} := m - q\Phi - q\phi$, where $\phi$ is now identified
with $\phi_{\sf R} = \phi[\text{tail}]$. These equations can be
simplified by inserting the background equation of motion, 
$(m-q\Phi) a_a = q\nabla_a \Phi$, on the right-hand side; taking into 
account the variation of $\Phi$ on the world line, the equation
implies $\dot{a}_a = (q/{\sf m}) \nabla_t \nabla_a \Phi 
+ 2(q/{\sf m})^2 \nabla_t \Phi \nabla_a \Phi$. The equations of motion
can also be simplified by inserting Eq.~(\ref{Ricci_scalar}) for the
background Ricci tensor. The end result is 
\begin{align} 
{\sf m} a_a &= q \nabla_a \Phi 
+ \frac{1}{6} q^2 \biggl( 5 + 4 \frac{q^2}{{\sf m}^2} 
- 11 \frac{{\sf m}^2}{q^2} \biggr) \nabla_t \Phi \nabla_a \Phi 
- \frac{1}{3} {\sf m} q \biggl( 5 
- \frac{q^2}{{\sf m}^2} \biggr)\nabla_t \nabla_a \Phi    
\nonumber \\ & \quad \mbox{}
- q h_{tt}[\text{tail}] \nabla_a \Phi 
- q h^t_{\ a}[\text{tail}] \nabla_t \Phi -
 q h^b_{\ a}[\text{tail}] \nabla_b \Phi 
+ q \phi_a [\text{tail}] 
- \frac{1}{2} {\sf m} \bigl( 2 h_{tat}[\text{tail}] 
- h_{tta}[\text{tail}] \bigr) 
\label{eom_scalar3} 
\end{align} 
and 
\begin{equation} 
\frac{d {\sf m}}{dt} = -q \nabla_t \Phi 
+ q^2 \biggl( \frac{11}{12} \nabla_\mu \Phi \nabla^\mu \Phi  
- \frac{1}{3} F + \frac{1}{2} F'' \biggr) 
+ {\sf m} q \Bigl( 2 \nabla_t \nabla_t \Phi + F' \Bigr) 
- q \phi_t[\text{tail}].  
\label{eom_scalar4} 
\end{equation} 

At this stage it is a simple matter to express the equations of
motion in covariant form. We have 
\begin{align} 
{\sf m} a^\mu &= \bigl( g^{\mu\nu} + u^\mu u^\nu \bigr) 
\biggl\{  q \nabla_\nu \Phi 
+ \frac{1}{6} q^2 \biggl( 5 + 4 \frac{q^2}{{\sf m}^2} 
- 11 \frac{{\sf m}^2}{q^2} \biggr) 
  u^\lambda \nabla_\lambda \Phi \nabla_\nu \Phi 
\nonumber \\ & \quad \mbox{}
- \frac{1}{3} {\sf m} q \biggl( 5 - \frac{q^2}{m^2} \biggr)
u^\lambda \nabla_{\lambda\nu} \Phi 
- q u^\lambda u^\rho h_{\lambda\rho}[\text{tail}] \nabla_\nu \Phi   
- q h^\lambda_{\ \nu}[\text{tail}] \nabla_\lambda \Phi 
\nonumber \\ & \quad \mbox{}
+ q \phi_\nu[\text{tail}] 
- \frac{1}{2} {\sf m} \bigl( 2 h_{\nu\lambda\rho}[\text{tail}] 
- h_{\lambda\rho\nu}[\text{tail}] \bigr) u^\lambda u^\rho \biggr\} 
\label{eom_scalar5} 
\end{align} 
and 
\begin{equation} 
\frac{d {\sf m}}{d\tau} = -q u^\mu \nabla_\mu \Phi 
+ q^2 \biggl( \frac{11}{12} \nabla^\mu \Phi \nabla_\mu \Phi  
- \frac{1}{3} F + \frac{1}{2} F'' \biggr) 
+ {\sf m} q \bigl( 2 u^\mu u^\nu \nabla_\mu \nabla_\nu \Phi 
+ F' \bigr) - q u^\mu \phi_\mu[\text{tail}].  
\label{eom_scalar6} 
\end{equation}  
The tail terms are still given by the equations displayed previously,
except that the dependence on $t^-$ can now be replaced by a
dependence on $\tau^- := \tau - 0^+$; the variable of integration
should then be replaced by $\tau'$.   

\subsection{Singular potentials} 
\label{subsec:sing_scalar} 

The equations of motion displayed in Eq.~(\ref{eom_scalar4}) and
(\ref{eom_scalar5}) are typically not the most useful starting point
to calculate the motion of a point particle. A more practical
formulation is based instead on the form 
\begin{equation} 
{\sf m} a^\mu = \bigl( g^{\mu\nu} + u^\mu u^\nu \bigr) \Bigl[ 
q \nabla_\nu \Phi 
- q u^\lambda u^\rho h^{\sf R}_{\lambda\rho} \nabla_\nu \Phi  
- q h^{{\sf R} \lambda}_{\ \ \nu} \nabla_\lambda \Phi 
+ q \nabla_\nu \phi^{\sf R}  
- \frac{1}{2} {\sf m} \bigl( 2 \nabla_\rho h^{\sf R}_{\nu\lambda} 
- \nabla_\nu h^{\sf R}_{\lambda\rho} \bigr) u^\lambda u^\rho \Bigr] 
\end{equation} 
and 
\begin{equation} 
\frac{d {\sf m}}{d\tau} = -q u^\mu \nabla_\mu \Phi 
- q u^\mu \nabla_\mu \phi^{\sf R}, 
\end{equation} 
in which the singular potentials $h^{\sf S}_{\alpha\beta}$ and 
$\phi^{\sf S}$ were explicitly removed from the retarded potentials
$h_{\alpha\beta}$ and $\phi$. This subtraction can be implemented by
formulating field equations for the regular potentials in terms of
extended effective sources, or by obtaining regularization
parameters when the self-force is computed as a sum over
spherical-harmonic modes.

The starting point of such computations is the singular potentials of
Eq.~(\ref{psi_sing_cov}), in which we insert
Eqs.~(\ref{UA_derivs}). With the results displayed in
Eqs.~(\ref{grav_scalar_tensorlist}), we have 
\begin{subequations} 
\begin{align} 
U^{\alpha\beta} &= g^{(\alpha}_{\ \hat{\alpha}}  
g^{\beta)}_{\ \hat{\beta}} \biggl\{ 
4{\sf m}  u^{\hat{\alpha}} u^{\hat{\beta}}   
\biggl( 1 + \frac{1}{12} R_{\hat{\mu}\hat{\nu}} 
\sigma^{\hat{\mu}} \sigma^{\hat{\nu}} \biggr) 
+ O(\rho^3) \biggr\} , \\   
\dot{U}^{\alpha\beta} &= g^{(\alpha}_{\ \hat{\alpha}} 
g^{\beta)}_{\ \hat{\beta}} \biggl\{ 
4{\sf m} \biggl( 2 a^{\hat{\alpha}} u^{\hat{\beta}}    
+ \frac{1}{6} u^{\hat{\alpha}} u^{\hat{\beta}} 
  R_{\hat{\mu}\hat{\nu}} u^{\hat{\mu}} \sigma^{\hat{\nu}}
+ R^{\hat{\alpha}}_{\ \hat{\gamma} \hat{\mu} \hat{\nu} } 
   u^{\hat{\beta}} u^{\hat{\gamma}} u^{\hat{\mu}} \sigma^{\hat{\nu}} 
\biggr) - 4 q \dot{\Phi} u^{\hat{\alpha}} u^{\hat{\beta}} 
+ O(\rho^2) \biggr\}, \\  
\ddot{U}^{\alpha\beta} &= g^{(\alpha}_{\ \hat{\alpha}} 
g^{\beta)}_{\ \hat{\beta}} \biggl\{ 
4{\sf m} \biggl( 2 \dot{a}^{\hat{\alpha}}
u^{\hat{\beta}} + 2 a^{\hat{\alpha}} a^{\hat{\beta}}    
+ \frac{1}{6} u^{\hat{\alpha}} u^{\hat{\beta}} 
  R_{\hat{\mu}\hat{\nu}} u^{\hat{\mu}} u^{\hat{\nu}} \biggr) 
-16 q \dot{\Phi} a^{\hat{\alpha}} u^{\hat{\beta}} 
- 4 q \ddot{\Phi} u^{\hat{\alpha}} u^{\hat{\beta}} 
+ O(\rho) \biggr\}, \\ 
V^{\alpha\beta} &= g^{(\alpha}_{\ \hat{\alpha}} 
g^{\beta)}_{\ \hat{\beta}} \biggl\{ 4{\sf m} \biggl( 
R^{\hat{\alpha}\ \hat{\beta}}_{\ \hat{\gamma}\ \hat{\delta}} 
u^{\hat{\gamma}} u^{\hat{\delta}} 
- u^{\hat{\alpha}} u^{\hat{\gamma}} \nabla^{\hat{\beta}} \Phi
\nabla_{\hat{\gamma}} \Phi 
+ \frac{1}{12} u^{\hat{\alpha}} u^{\hat{\beta}} R \biggr) 
- 2 q \nabla^{\hat{\alpha}} \nabla^{\hat{\beta}} \Phi 
+ O(\rho) \biggr\} 
\end{align} 
\end{subequations} 
and 
\begin{subequations} 
\begin{align} 
U^\cdot &= q \biggl( 1 
+ \frac{1}{12} R_{\hat{\mu}\hat{\nu}} \sigma^{\hat{\mu}}
\sigma^{\hat{\nu}} \biggr) + O(\rho^3), \\ 
\dot{U}^\cdot &= \frac{1}{6} q  
R_{\hat{\mu}\hat{\nu}} u^{\hat{\mu}} \sigma^{\hat{\nu}} 
+ O(\rho^2), \\ 
\ddot{U}^\cdot &= \frac{1}{6} q  
R_{\hat{\mu}\hat{\nu}} u^{\hat{\mu}} u^{\hat{\nu}} 
+ O(\rho), \\
V^\cdot &= -\frac{1}{2} q \biggl( 2 \nabla_{\hat{\mu}} \Phi 
\nabla^{\hat{\mu}} \Phi + F'' - \frac{1}{6} R \biggr) 
- 2{\sf m} \biggl( u^{\hat{\alpha}} u^{\hat{\beta}} 
\nabla_{\hat{\alpha}} \nabla_{\hat{\beta}} \Phi
+ \frac{1}{2} F' \biggr) + O(\rho). 
\end{align} 
\end{subequations}  

\section{Self-force in electrovac spacetimes} 
\label{sec:em_sf} 

In this section we compute the self-force acting on a particle of mass
$m$ and electric charge $e$ in an electrovac spacetime; the metric of
the background spacetime is $g_{\alpha\beta}$, and the background
electromagnetic field is $F_{\alpha\beta}$. 

\subsection{Equations of motion} 

On a formal level, the motion of the particle is governed by the
Lorentz-force equation   
\begin{equation} 
m \frac{ {\sf D} {\sf u}^\mu }{d {\sf t}} 
= e {\sf F}^\mu_{\ \nu} {\sf u}^{\nu},  
\end{equation} 
in which $\sf t$ is proper time in the perturbed spacetime, 
${\sf u}^\mu = dz^\mu/d{\sf t}$, and ${\sf D}/d{\sf t}$ indicates
covariant differentiation (in the perturbed spacetime) along the world 
line. Substitution of ${\sf g}_{\alpha\beta} = g_{\alpha\beta} 
+ h_{\alpha\beta}$, ${\sf F}_{\alpha\beta} = F_{\alpha\beta} 
+ f_{\alpha\beta}$, and linearization with respect to all
perturbations produce   
\begin{equation} 
m a^\mu = e F^\mu_{\ \nu} u^\nu 
- \frac{1}{2} e u^\lambda u^\rho h_{\lambda\rho} F^\mu_{\ \nu} u^\nu 
- e \bigl( g^{\mu\nu} + u^\mu u^\nu \bigr) h_{\nu\lambda} 
F^\lambda_{\ \rho} u^\rho + e f^{\mu}_{\ \nu} u^\nu 
- \frac{1}{2} m \bigl( g^{\mu\nu} + u^\mu u^\nu \bigr) 
\bigl( 2 \nabla_\rho h_{\nu\lambda} 
- \nabla_\nu h_{\lambda\rho} \bigr) u^\lambda u^\rho,  
\label{sf_grav_em1} 
\end{equation} 
in which all quantities now refer to the background spacetime. To make
sense of these equations we continue to follow the Detweiler-Whiting
prescription to remove the singular piece of all potentials. The
equations of motion continue to be given by Eq.~(\ref{sf_grav_em1}),
but with the regular potentials standing in for the retarded
potentials.  

In Fermi coordinates we have 
\begin{equation} 
m a_a = e F_{at} - \frac{1}{2} e h_{tt} F_{at} - e h_{ab} F^b_{\ t} 
+ e f_{at} - \frac{1}{2} m \bigl( 2 \nabla_t h_{ta} 
- \nabla_a h_{tt} \bigr), 
\label{sf_grav_em2} 
\end{equation} 
where we suppress the label ``${\sf R}$'' on the potentials.

\subsection{Regular potentials} 

Importing $M^A_{\ B\mu}$ and $N^A_{\ B}$ from
Sec.~\ref{sec:electrovac} and inserting them within
Eqs.~(\ref{U_taylor2}) and (\ref{V_taylor2}) reveals that the tensors
that appear in the Hadamard Green's function are given explicitly by  
\begin{subequations} 
\label{grav_em_tensorlist} 
\begin{align} 
U^{\alpha\beta}_{\ \ \ |\gamma\mu} &=  
8 \delta^{(\alpha}_{\ [\mu} F^{\beta)}_{\ \gamma]} 
- 2 g^{\alpha\beta} F_{\mu\gamma}, \\ 
U^\alpha_{\ \ |\beta\gamma\mu} &= 
\frac{1}{2} \delta^\alpha_{\ (\beta} F_{\gamma)\mu} 
- \frac{1}{4} g_{\beta\gamma} F^\alpha_{\ \mu}, \\ 
U^{\alpha\beta}_{\ \ \ |\gamma\delta\mu\nu} &= 
\delta^{(\alpha}_{\ \mu} F^{\beta)}_{\ (\gamma} F_{\delta)\nu} 
+ \delta^{(\alpha}_{\ \nu} F^{\beta)}_{\ (\gamma} F_{\delta)\mu} 
- \delta^{(\alpha}_{\ (\gamma} F^{\beta)}_{\ \mu} F_{\delta)\nu} 
- \delta^{(\alpha}_{\ (\gamma} F^{\beta)}_{\ \nu} F_{\delta)\mu}  
- g^{\alpha\beta} F_{\mu(\gamma} F_{\delta)\nu} 
- \frac{1}{2} g_{\gamma\delta} 
  \delta^{(\alpha}_{\ \mu} F^{\beta)}_{\ \lambda} F^\lambda_{\ \nu}  
\nonumber \\ & \quad \mbox{} 
- \frac{1}{2} g_{\gamma\delta} 
  \delta^{(\alpha}_{\ \nu} F^{\beta)}_{\ \lambda} F^\lambda_{\ \mu}  
+ g_{\gamma\delta} F^{(\alpha}_{\ \ \mu} F^{\beta)}_{\ \nu} 
+ \frac{1}{2} g^{\alpha\beta} g_{\gamma\delta} 
  F_{\mu\lambda} F^{\lambda}_{\ \nu} 
+ \frac{1}{6} \delta^{(\alpha}_{\ \gamma} \delta^{\beta)}_{\ \delta} 
  R_{\mu\nu}, \\ 
U^{\alpha\beta}_{\ \ \ |\gamma\mu\nu} &= 
- 4 \delta^{(\alpha}_{\ (\mu} \nabla_{\nu)} F^{\beta)}_{\ \gamma} 
+ 4 \delta^{(\alpha}_{\ \gamma} \nabla_{(\mu} F^{\beta)}_{\ \nu)} 
+ 2 g^{\alpha\beta} \nabla_{(\mu} F_{\nu)\gamma}, \\ 
U^\alpha_{\ |\beta\gamma\mu\nu} &= -\frac{1}{4} \Bigl( 
\delta^\alpha_{\ (\beta} \nabla_\mu F_{\gamma)\nu} 
+ \delta^\alpha_{\ (\beta} \nabla_\nu F_{\gamma)\mu} 
- g_{\beta\gamma} \nabla_{(\mu} F^\alpha_{\ \nu)} \Bigr), \\
U^\alpha_{\ \ |\beta\mu\nu} &= 
- \delta^{\alpha}_{\ (\mu} F_{\nu)\lambda} F^\lambda_{\ \beta} 
+ \delta^{\alpha}_{\ \beta} F_{\mu\lambda} F^{\lambda}_{\ \nu} 
+ \frac{1}{6} \delta^\alpha_{\ \beta} R_{\mu\nu}, \\ 
V^{\alpha\beta}_{\ \ \ |\gamma\delta} &= 
R^{\alpha\ \beta}_{\ (\gamma\ \delta)}  
+ \frac{1}{12} \delta^\alpha_{\ (\gamma} \delta^\beta_{\ \delta)} R 
- F^\alpha_{\ (\gamma} F^\beta_{\ \delta)} 
- \delta^{(\alpha}_{\ (\gamma} F^{\beta)\lambda} F_{\delta)\lambda} 
+ \frac{1}{2} g_{\alpha\beta} F^\lambda_{\ (\gamma} F_{\delta)\lambda} 
+ \frac{1}{4} g^{\alpha\beta} g_{\gamma\delta} 
  F_{\lambda\rho} F^{\lambda\rho}, \\ 
V^{\alpha\beta}_{\ \ \ |\gamma} &= 
-2 \nabla^{(\alpha} F^{\beta)}_{\ \gamma}, \\ 
V^\alpha_{\ \ |\beta\gamma} &= 
\frac{1}{2} \nabla_{(\beta} F^\alpha_{\ \gamma)}, \\ 
V^\alpha_{\ |\beta} &= -\frac{1}{2} R^\alpha_{\ \beta} 
+ \frac{1}{12} \delta^{\alpha}_{\ \beta} R 
- \frac{1}{2} F^{\alpha \lambda} F_{\beta\lambda} 
+ \frac{1}{2} \delta^\alpha_{\ \beta} 
  F_{\lambda\rho} F^{\lambda\rho},  
\end{align} 
\end{subequations}  
where we omit the primes on indices to keep the notation uncluttered.  

Making the substitutions in Eq.~(\ref{psi_fermi_b}), we obtain 
\begin{subequations} 
\begin{align} 
\gamma^{tt}_{\sf R} &= \frac{2}{3} m \bigl( 2 \dot{a}_c + R_{tc}
\bigr) x^c + e \bigl( -\nabla_t F_{ct} + 2 a^d F_{cd} \bigr) x^c 
+ \gamma^{tt}[\text{tail}] + O(s^2), \\ 
\gamma^{ta}_{\sf R} &= -4m a^a + m \bigl( 8 a^a a_c 
- 2 R^a_{\ tct} + F^a_{\ t} F_{tc} + F^{a e} F_{ec} 
+ \delta^a_{\ c} F_{et} F^e_{\ t} \bigr) x^c 
\nonumber \\ & \quad \mbox{} 
+ e \bigl( -\nabla_t F^a_{\ c} 
- 2 \delta^a_{\ c} a^e F_{te} + 2 a^a F_{tc} \bigr) x^c  
+ \gamma^{ta}[\text{tail}] + O(s^2), \\ 
\gamma^{ab}_{\sf R} &= m \bigl( 
2 \delta^{(a}_{\ c} F^{b)}_{\ e} F^e_{\ t} 
- 4 F^{(a}_{\ \ t} F^{b)}_{\ c} 
+ 2 \delta^{ab} F_{et} F^e_{\ c} \bigr) x^c 
+ e \bigl( 2 \delta^{(a}_{\ c} \nabla_t F^{b)}_{\ t} 
- \delta^{ab} \nabla_t F_{ct} 
\nonumber \\ & \quad \mbox{} 
+ 4 a^e \delta^{(a}_{\ c} F^{b)}_{\ e} 
- 4 a^{(a} F^{b)}_{\ c} 
- 2 \delta^{ab} a^e F_{ce} \bigr) x^c   
+ \gamma^{ab}[\text{tail}] + O(s^2) 
\end{align} 
\end{subequations} 
for the gravitational potentials, and 
\begin{subequations} 
\begin{align} 
b^t_{\sf R} &= \frac{1}{6} e \bigl( 2 \dot{a}_c + 3 F_{ce} F^e_{\ t} 
+ R_{tc} \bigr) x^c 
- \frac{1}{2} m \bigl( \nabla_t F_{ct} + 4 a^e F_{ce} \bigr) x^c 
+ b^t[\text{tail}] + O(s^2), \\ 
b^a_{\sf R} &= -e a^a - m F^a_{\ t} + \frac{1}{2} e \bigl( 2 a_c a^a 
- R^a_{\ tct} + \delta^a_{\ c} F_{et} F^e_{\ t} \bigr) x^c 
\nonumber \\ & \quad \mbox{} 
+ \frac{1}{2} m \bigl( 2 a_c F^a_{\ t} - 4 a^a F_{ct} 
+ \nabla_t F^a_{\ c} - \nabla_c F^a_{\ t} \bigr) x^c 
+ b^a[\text{tail}] + O(s^2)
\end{align} 
\end{subequations} 
for the electromagnetic potentials. The tail terms are 
\begin{equation} 
\gamma^{\alpha\beta}[\text{tail}] := 
4m \int_{-\infty}^{t^-} G^{\alpha\beta}_{\ \ \ |\mu\nu}(x,z) 
u^\mu u^\nu\, d\tau 
+ e \int_{-\infty}^{t^-} G^{\alpha\beta}_{\ \ \ |\mu}(x,z) 
u^\mu\, d\tau
\end{equation} 
and 
\begin{equation} 
b^\alpha[\text{tail}] := 
4m \int_{-\infty}^{t^-} G^{\alpha}_{\ |\mu\nu}(x,z) 
u^\mu u^\nu\, d\tau 
+ e \int_{-\infty}^{t^-} G^{\alpha}_{\ |\mu}(x,z)
u^\mu\, d\tau.  
\label{em_b_tail} 
\end{equation}
After trace reversal and lowering the indices using the metric 
$g_{tt} = -1 - 2a_c x^c + O(s^2)$, $g_{ta} = O(s^2)$, 
$g_{ab} = \delta_{ab} + O(s^2)$, the gravitational potentials become 
\begin{subequations} 
\begin{align} 
h_{tt}^{\sf R} &= \frac{1}{3} m \bigl( 2 \dot{a}_c + R_{tc}
\bigr) x^c + e \bigl( -\nabla_t F_{ct} + 2 a^d F_{cd} \bigr) x^c 
+ h_{tt}[\text{tail}] + O(s^2), \\ 
h_{ta}^{\sf R} &= 4m a_a + m \bigl( 2 R_{atct} 
+ F_{at} F_{ct} - F_{a e} F^e_{\ c} 
- \delta_{ac} F_{et} F^e_{\ t} \bigr) x^c 
\nonumber \\ & \quad \mbox{} 
+ e \bigl( \nabla_t F_{ac} 
- 2 \delta_{ac} a^e F_{et} + 2 a_a F_{ct} \bigr) x^c  
+ h_{ta}[\text{tail}] + O(s^2), \\ 
h_{ab}^{\sf R} &= \frac{1}{3} m \delta_{ab} \bigl( 2\dot{a}_c 
+ R_{tc} \bigr) x^c 
+ m \bigl( 2 \delta_{c(a} F_{b)e} F^e_{\ t} 
+ 4 F_{t(a} F_{b)c} 
+ 2 \delta_{ab} F_{et} F^e_{\ c} \bigr) x^c 
\nonumber \\ & \quad \mbox{} 
+ e \bigl( 2 \delta_{c(a} \nabla_t F_{b) t} 
+ \delta_{ab} \nabla_t F_{tc} 
+ 4 a^e \delta_{c(a} F_{b)e} 
- 4 a_{(a} F_{b) c} 
- 2 \delta_{ab} a^e F_{ce} \bigr) x^c   
+ h_{ab}[\text{tail}] + O(s^2),  
\end{align} 
\end{subequations} 
with tail terms now given by 
\begin{equation} 
h_{\alpha\beta}[\text{tail}] := 
4m \int_{-\infty}^{t^-} \bar{G}_{\alpha\beta|\mu\nu}(x,z) 
u^\mu u^\nu\, d\tau 
+ e \int_{-\infty}^{t^-} \bar{G}_{\alpha\beta|\mu}(x,z)
u^\mu \, d\tau,
\label{em_h_tail} 
\end{equation} 
where the overbar indicates the operation of trace reversal. The
electromagnetic potentials become  
\begin{subequations} 
\begin{align} 
b_t^{\sf R} &= -\frac{1}{6} e \bigl( 2 \dot{a}_c + 3 F_{ce} F^e_{\ t} 
+ R_{tc} \bigr) x^c 
+ \frac{1}{2} m \bigl( \nabla_t F_{ct} + 4 a^d F_{cd} \bigr) x^c 
+ b_t[\text{tail}] + O(s^2), \\ 
b_a^{\sf R} &= -e a_a - m F_{at} + \frac{1}{2} e \bigl( 2 a_c a_a 
- R_{atct} + \delta_{ac} F_{et} F^e_{\ t} \bigr) x^c 
\nonumber \\ & \quad \mbox{} 
+ \frac{1}{2} m \bigl( 2 a_c F_{a t} - 4 a_a F_{ct} 
+ \nabla_t F_{ac} - \nabla_c F_{a t} \bigr) x^c 
+ b_a[\text{tail}] + O(s^2)
\end{align} 
\end{subequations} 
after lowering the indices.  

The relevant covariant derivatives of the potentials are 
\begin{subequations} 
\begin{align} 
\nabla_t h_{ta} &= 4m \bigl( \dot{a}_a + \bar{V}_{ta|tt} \bigr) 
+ e \bar{V}_{ta|t} + h_{tat}[\text{tail}] + O(s), \\ 
\nabla_a h_{tt} &= \frac{1}{3} m \bigl( 2\dot{a}_a + R_{ta}\bigr)  
+ e \bigl( -\nabla_t F_{at} + 2 a^c F_{ac} \bigr) 
+ h_{tta}[\text{tail}] + O(s), \\ 
\nabla_t b_a &= e \bigl( -\dot{a}_a + V_{a|t} \bigr) 
+ m \bigl( -\nabla_t F_{at} - a^c F_{ac} + 4 V_{a|tt} \bigr) 
+ b_{at}[\text{tail}] + O(s), \\ 
\nabla_a b_t &= -\frac{1}{6} e \bigl( 2 \dot{a}_a 
+ 3 F_{ac} F^c_{\ t} + R_{ta} \bigr) 
+ \frac{1}{2} m \bigl( \nabla_t F_{at} + 4 a^c F_{ac} \bigr) 
+ b_{ta}[\text{tail}] + O(s), 
\end{align} 
\end{subequations}  
where 
\begin{subequations}
\label{em_grad_tail} 
\begin{align} 
h_{\alpha\beta\gamma}[\text{tail}] &:= 
4m \int_{-\infty}^{t^-} \nabla_\gamma 
\bar{G}_{\alpha\beta|\mu\nu}(x,z) u^\mu u^\nu\, d\tau 
+ e \int_{-\infty}^{t^-} \nabla_\gamma 
\bar{G}_{\alpha\beta|\mu}(x,z) u^\mu\, d\tau, \\
b_{\alpha\beta}[\text{tail}] &:= 
4m \int_{-\infty}^{t^-} \nabla_\beta 
G_{\alpha|\mu\nu}(x,z) u^\mu u^\nu\, d\tau 
+ e \int_{-\infty}^{t^-} \nabla_\beta  
G_{\alpha|\mu}(x,z) u^\mu\, d\tau. 
\end{align} 
\end{subequations} 

\subsection{Explicit form of the equations of motion} 

Making the substitutions in Eq.~(\ref{sf_grav_em2}), we obtain 
\begin{align} 
m a_a &= e F_{at} + m^2 \biggl( -\frac{11}{3} \dot{a}_a 
+ \frac{1}{6} R_{ta} + 2 F_{ac} F^c_{\ t} \biggr) 
+ e^2 \biggl( \frac{2}{3} \dot{a}_a + \frac{1}{3} R_{ta} 
- F_{ac} F^c_{\ t} \biggr) 
+ 4 m e a^c F_{ac} 
\nonumber \\ & \quad \mbox{} 
- \frac{1}{2} e h_{tt}[\text{tail}] F_{at} 
- e h_{ab}[\text{tail}] F^b_{\ t} 
+ e \bigl( b_{ta}[\text{tail}] - b_{at}[\text{tail}] \bigr) 
- \frac{1}{2} m \bigl( 2 h_{tat}[\text{tail}] 
- h_{tta}[\text{tail}] \bigr).  
\label{eom_em1} 
\end{align} 
This is simplified by inserting the background equation of motion,
$m a_a = e F_{at}$, on the right-hand side; the equation implies 
$\dot{a}_a = (e/m) \nabla_t F_{at} + (e/m)^2 F_{ac} F^c_{\ t}$. It is
also simplified by inserting $R_{ta} = -2 F_{ac} F^c_{\ t}$ for the
background Ricci tensor. The end result is 
\begin{align} 
m a_a &= e F_{at} 
- \frac{1}{3} m e \biggl( 11 -  2 \frac{e^2}{m^2} \biggr) \nabla_t F_{at} 
- \frac{1}{3} e^2 \biggl( 4 - 2 \frac{e^2}{m^2} - 5 \frac{m^2}{e^2}
   \biggr) F_{ac} F^c_{\ t} 
\nonumber \\ & \quad \mbox{} 
- \frac{1}{2} e h_{tt}[\text{tail}] F_{at} 
- e h_{ab}[\text{tail}] F^b_{\ t} 
+ e \bigl( b_{ta}[\text{tail}] - b_{at}[\text{tail}] \bigr) 
- \frac{1}{2} m \bigl( 2 h_{tat}[\text{tail}] 
- h_{tta}[\text{tail}] \bigr).  
\label{eom_em2} 
\end{align} 
The covariant form of the equations of motion is 
\begin{align} 
m a^\mu &= e F^\mu_{\ \nu} u^\nu 
+ \bigl( g^{\mu\nu} + u^\mu u^\nu \bigr) \Biggl\{ 
- \frac{1}{3} m e \biggl( 11 -  2 \frac{e^2}{m^2} \biggr)  
  u^\lambda u^\rho \nabla_\rho F_{\nu\lambda} 
- \frac{1}{3} e^2 \biggl( 4 - 2 \frac{e^2}{m^2} - 5 \frac{m^2}{e^2}
   \biggr) F_{\nu\lambda} F^\lambda_{\ \rho} u^\rho 
\nonumber \\ & \quad \mbox{} 
- \frac{1}{2} e u^\lambda u^\rho h_{\lambda\rho}[\text{tail}] 
   F_{\nu\tau} u^\tau 
- e h_{\nu\lambda}[\text{tail}] F^\lambda_{\ \rho} u^\rho 
+ e \bigl( b_{\lambda\nu}[\text{tail}] - b_{\nu\lambda}[\text{tail}]
   \bigr) u^\lambda  
\nonumber \\ & \quad \mbox{} 
- \frac{1}{2} m \bigl( 2 h_{\nu\lambda\rho}[\text{tail}] 
- h_{\lambda\rho\nu}[\text{tail}] \bigr) u^\lambda u^\rho \Biggr\}. 
\label{eom_em3} 
\end{align} 
The tail terms are still given by the equations displayed previously, 
except that the dependence on $t^-$ can now be replaced by a
dependence on $\tau^- := \tau - 0^+$; the variable of integration
should then be replaced by $\tau'$.   

\subsection{Singular potentials} 
\label{subsec:sing_em} 

The equation of motion displayed in Eq.~(\ref{eom_em3}) is typically
not the most useful starting point to calculate the motion of a point
particle. A more practical formulation is based instead on the form 
\begin{equation} 
m a^\mu = e F^\mu_{\ \nu} u^\nu 
- \frac{1}{2} e u^\lambda u^\rho h^{\sf R}_{\lambda\rho} 
   F^\mu_{\ \nu} u^\nu  
- e \bigl( g^{\mu\nu} + u^\mu u^\nu \bigr) h^{\sf R}_{\nu\lambda}  
F^\lambda_{\ \rho} u^\rho + e f^{\mu}_{{\sf R}\, \nu} u^\nu 
- \frac{1}{2} m \bigl( 2 \nabla_\rho h^{\sf R}_{\nu\lambda} 
- \nabla_\nu h^{\sf R}_{\lambda\rho} \bigr) u^\lambda u^\rho,   
\end{equation} 
in which the singular potentials $h^{\sf S}_{\alpha\beta}$ and 
$b^{\sf S}_\alpha$ were explicitly removed from the retarded
potentials $h_{\alpha\beta}$ and $b_\alpha$. This subtraction can be
implemented by formulating field equations for the regular potentials
in terms of extended effective sources, or by obtaining
regularization parameters when the self-force is computed as a sum
over spherical-harmonic modes. 

The starting point of such computations is the singular potentials of
Eq.~(\ref{psi_sing_cov}), in which we insert
Eqs.~(\ref{UA_derivs}). The explicit expressions are 
\begin{subequations} 
\begin{align} 
U^{\alpha\beta} &= g^{(\alpha}_{\ \hat{\alpha}} 
g^{\beta)}_{\ \hat{\beta}} \biggl\{ 
4m \biggl( u^{\hat{\alpha}} u^{\hat{\beta}}  
+ \frac{1}{2} U^{\hat{\alpha}\hat{\beta}}_{\ \ \ 
                 |\hat{\gamma}\hat{\delta}\hat{\mu}\hat{\nu}} 
  u^{\hat{\gamma}} u^{\hat{\delta}} 
  \sigma^{\hat{\mu}} \sigma^{\hat{\nu}} \biggr) 
+ e \biggl( 
  U^{\hat{\alpha}\hat{\beta}}_{\ \ \ |\hat{\gamma}\hat{\mu}} 
  u^{\hat{\gamma}} \sigma^{\hat{\mu}}
+ \frac{1}{2} 
  U^{\hat{\alpha}\hat{\beta}}_{\ \ \ |\hat{\gamma}\hat{\mu}\hat{\nu}}  
  u^{\hat{\gamma}} \sigma^{\hat{\mu}} \sigma^{\hat{\nu}}\biggr) 
+ O(\rho^3) \biggr\}, \\ 
\dot{U}^{\alpha\beta} &= g^{(\alpha}_{\ \hat{\alpha}} 
g^{\beta)}_{\ \hat{\beta}} \biggl\{ 
4m \biggl( 
2 a^{\hat{\alpha}} u^{\hat{\beta}}   
+ U^{\hat{\alpha}\hat{\beta}}_{\ \ \
                 |\hat{\gamma}\hat{\delta}\hat{\mu}\hat{\nu}} 
  u^{\hat{\gamma}} u^{\hat{\delta}} u^{\hat{\mu}} \sigma^{\hat{\nu}} 
+ R^{\hat{\alpha}}_{\ \hat{\gamma}\hat{\mu}\hat{\nu}} 
    u^{\hat{\beta}} u^{\hat{\gamma}} u^{\hat{\mu}} \sigma^{\hat{\nu}} 
\biggr) 
\nonumber \\ & \quad \mbox{} 
+ e \biggl(  
U^{\hat{\alpha}\hat{\beta}}_{\ \ \ |\hat{\gamma}\hat{\mu}} 
  u^{\hat{\gamma}} u^{\hat{\mu}}
+ \dot{U}^{\hat{\alpha}\hat{\beta}}_{\ \ \ |\hat{\gamma}\hat{\mu}} 
  u^{\hat{\gamma}} \sigma^{\hat{\mu}}
+ U^{\hat{\alpha}\hat{\beta}}_{\ \ \ |\hat{\gamma}\hat{\mu}} 
  a^{\hat{\gamma}} \sigma^{\hat{\mu}}
+ U^{\hat{\alpha}\hat{\beta}}_{\ \ \ |\hat{\gamma}\hat{\mu}\hat{\nu}}   
  u^{\hat{\gamma}} u^{\hat{\mu}} \sigma^{\hat{\nu}}\biggr)  
+ O(\rho^2) \biggr\}, \\ 
\ddot{U}^{\alpha\beta} &= g^{(\alpha}_{\ \hat{\alpha}} 
g^{\beta)}_{\ \hat{\beta}} \biggl\{ 
4m \biggl( 
2 \dot{a}^{\hat{\alpha}} u^{\hat{\beta}}  
+ 2 a^{\hat{\alpha}} a^{\hat{\beta}}  
+ U^{\hat{\alpha}\hat{\beta}}_{\ \ \
                 |\hat{\gamma}\hat{\delta}\hat{\mu}\hat{\nu}} 
  u^{\hat{\gamma}} u^{\hat{\delta}} u^{\hat{\mu}} u^{\hat{\nu}} 
\biggr) 
\nonumber \\ & \quad \mbox{} 
+ e \biggl(  
2 U^{\hat{\alpha}\hat{\beta}}_{\ \ \ |\hat{\gamma}\hat{\mu}} 
  a^{\hat{\gamma}} u^{\hat{\mu}}
+ 2 \dot{U}^{\hat{\alpha}\hat{\beta}}_{\ \ \ |\hat{\gamma}\hat{\mu}} 
  u^{\hat{\gamma}} u^{\hat{\mu}}
+ U^{\hat{\alpha}\hat{\beta}}_{\ \ \ |\hat{\gamma}\hat{\mu}} 
   u^{\hat{\gamma}} a^{\hat{\mu}}
+ U^{\hat{\alpha}\hat{\beta}}_{\ \ \ |\hat{\gamma}\hat{\mu}\hat{\nu}}   
  u^{\hat{\gamma}} u^{\hat{\mu}} u^{\hat{\nu}}\biggr)  
+ O(\rho) \biggr\}, \\ 
V^{\alpha\beta} &= g^{(\alpha}_{\ \hat{\alpha}} 
g^{\beta)}_{\ \hat{\beta}} \biggl\{ 
4m V^{\hat{\alpha}\hat{\beta}}_{\ \ \ |\hat{\gamma}\hat{\delta}}  
  u^{\hat{\gamma}} u^{\hat{\delta}} 
+ e V^{\hat{\alpha}\hat{\beta}}_{\ \ \ |\hat{\gamma}}   
  u^{\hat{\gamma}} + O(\rho) \biggr\},
\end{align} 
\end{subequations} 
and 
\begin{subequations} 
\begin{align} 
U^\alpha &= g^{\alpha}_{\ \hat{\alpha}} \biggl\{ 
e \biggl( u^{\hat{\alpha}}   
+ \frac{1}{2} U^{\hat{\alpha}}_{\ \
                          |\hat{\beta}\hat{\mu}\hat{\nu}}  
  u^{\hat{\beta}} \sigma^{\hat{\mu}} \sigma^{\hat{\nu}} \biggr)  
+ 4m \biggl( 
  U^{\hat{\alpha}}_{\ \ |\hat{\beta}\hat{\gamma}\hat{\mu}} 
  u^{\hat{\beta}} u^{\hat{\gamma}} \sigma^{\hat{\mu}}
+ \frac{1}{2} 
  U^{\hat{\alpha}}_{\ \ |\hat{\beta}\hat{\gamma}\hat{\mu}\hat{\nu}}  
  u^{\hat{\beta}} u^{\hat{\gamma}} 
  \sigma^{\hat{\mu}} \sigma^{\hat{\nu}} \biggr) 
+ O(\rho^3) \biggr\}, \\ 
\dot{U}^\alpha &= g^{\alpha}_{\ \hat{\alpha}} \biggl\{ 
e \biggl( a^{\hat{\alpha}}   
+ U^{\hat{\alpha}}_{\ \ |\hat{\beta}\hat{\mu}\hat{\nu}}  
  u^{\hat{\beta}} u^{\hat{\mu}} \sigma^{\hat{\nu}}
+ \frac{1}{2} R^{\hat{\alpha}}_{\ \hat{\beta}\hat{\mu}\hat{\nu}} 
  u^{\hat{\beta}} u^{\hat{\mu}} \sigma^{\hat{\nu}} \biggr) 
\nonumber \\ & \quad \mbox{} 
+ 4m \biggl( 
  U^{\hat{\alpha}}_{\ \ |\hat{\beta}\hat{\gamma}\hat{\mu}} 
  u^{\hat{\beta}} u^{\hat{\gamma}} u^{\hat{\mu}}
+ \dot{U}^{\hat{\alpha}}_{\ \ |\hat{\beta}\hat{\gamma}\hat{\mu}} 
  u^{\hat{\beta}} u^{\hat{\gamma}} \sigma^{\hat{\mu}} 
+ 2 U^{\hat{\alpha}}_{\ \ |\hat{\beta}\hat{\gamma}\hat{\mu}} 
  a^{\hat{\beta}} u^{\hat{\gamma}} \sigma^{\hat{\mu}} 
+ U^{\hat{\alpha}}_{\ \ |\hat{\beta}\hat{\gamma}\hat{\mu}\hat{\nu}}  
  u^{\hat{\beta}} u^{\hat{\gamma}} u^{\hat{\mu}} \sigma^{\hat{\nu}} 
\biggr) + O(\rho^2) \biggr\}, \\
\ddot{U}^\alpha &= g^{\alpha}_{\ \hat{\alpha}} \biggl\{  
e \biggl( \dot{a}^{\hat{\alpha}}   
+ U^{\hat{\alpha}}_{\ \ |\hat{\beta}\hat{\mu}\hat{\nu}}  
  u^{\hat{\gamma}} u^{\hat{\mu}} u^{\hat{\nu}} \biggr) 
\nonumber \\ & \quad \mbox{} 
+ 4m \biggl( 
4 U^{\hat{\alpha}}_{\ \ |\hat{\beta}\hat{\gamma}\hat{\mu}} 
  a^{\hat{\beta}} u^{\hat{\gamma}} u^{\hat{\mu}}
+ 2 \dot{U}^{\hat{\alpha}}_{\ \ |\hat{\beta}\hat{\gamma}\hat{\mu}} 
  u^{\hat{\beta}} u^{\hat{\gamma}} u^{\hat{\mu}} 
+ U^{\hat{\alpha}}_{\ \ |\hat{\beta}\hat{\gamma}\hat{\mu}} 
  u^{\hat{\beta}} u^{\hat{\gamma}} a^{\hat{\mu}} 
+ U^{\hat{\alpha}}_{\ \ |\hat{\beta}\hat{\gamma}\hat{\mu}\hat{\nu}}  
  u^{\hat{\beta}} u^{\hat{\gamma}} u^{\hat{\mu}} u^{\hat{\nu}} 
\biggr) + O(\rho) \biggr\}, \\ 
V^\alpha &= g^{\alpha}_{\ \hat{\alpha}} \biggl\{ 
e V^{\hat{\alpha}}_{\ \ |\hat{\beta}} u^{\hat{\beta}}
+ 4m V^{\hat{\alpha}}_{\ \ |\hat{\beta}\hat{\gamma}}  
  u^{\hat{\beta}} u^{\hat{\gamma}} + O(\rho) \biggr\}. 
\end{align} 
\end{subequations} 
The various tensors that appear in these expressions were displayed in 
Eqs.~(\ref{grav_em_tensorlist}).  

\begin{acknowledgments} 
We thank Thomas Linz, John Friedman, and Alan Wiseman for sharing
their results. This work was supported by the Natural Sciences and
Engineering Research Council of Canada.      
\end{acknowledgments}    

\bibliography{../bib/master} 
\end{document}